\pdfoutput=1
\documentclass[conference]{IEEEtran}

\usepackage{cite}
\usepackage{amsmath,amssymb,amsfonts}
\usepackage{graphicx}
\usepackage{textcomp}
\usepackage{xcolor}
\usepackage{color, colortbl}
\usepackage{algorithm}
\usepackage{algpseudocode}
\usepackage{booktabs} 
\usepackage[export]{adjustbox}
\usepackage{xspace}
\usepackage[hyphens]{url}
\usepackage{color, verbatim}
\usepackage{mathrsfs}
\usepackage{amsthm}
\usepackage{mdframed}
\usepackage{threeparttable}
\newcommand{\squishcount}{
   \begin{list}{\arabic{enumi})}
     { \usecounter{enumi}
       \setlength{\itemindent}{0em}
       \setlength{\parskip}{0pt}
      \setlength{\itemsep}{0pt}      \setlength{\parsep}{1pt}
      \setlength{\topsep}{1pt}       \setlength{\partopsep}{0pt}
      \setlength{\leftmargin}{1.5em} \setlength{\labelwidth}{10em}
      \setlength{\labelsep}{0.5em} } }

\usepackage{array}
\newcommand{\PreserveBackslash}[1]{\let\temp=\\#1\let\\=\temp}
\newcolumntype{C}[1]{>{\PreserveBackslash\centering}p{#1}}
\newcolumntype{R}[1]{>{\PreserveBackslash\raggedleft}p{#1}}
\newcolumntype{L}[1]{>{\PreserveBackslash\raggedright}p{#1}}

\newcommand{\countend}{
  \end{list}}

\usepackage[normalem]{ulem}

\newcommand{\brianedit}[1]{{#1}}

\newcommand{\rahmedit}[1]{{#1}}
\newcommand{\athina}[1]{{#1}}

\newcommand{\jvedit}[1]{{#1}}

\newcommand{\majorrevision}[1]{{#1}}

\newcommand{\eg}{\hbox{{\em e.g.,}}\xspace}
\newcommand{\ie}{\hbox{{\em i.e.,}}\xspace}
\newcommand{\etal}{\textit{et al}.\xspace}

\newcommand{\etc}{\hbox{etc.}\xspace}
\newcommand{\code}[1]{\text{\tt #1}}
\newcommand{\myparagraph}[1]{\vspace{.2em}\noindent {\bf #1}\hspace{.1em}}
\newcommand{\mysubparagraph}[1]{\vspace{.2em}\noindent {\it #1}\hspace{.1em}}

\newcommand{\mycomment}[1]{}

\newcommand{\appdx}[1]{#1}
\newcommand{\techreport}[1]{#1}
\newcommand{\techreportremove}[1]{}

\newcommand{\tool}[0]{\textsc{PingPong}\xspace}
\newcommand{\toolping}[0]{\textsc{Ping}\xspace}
\newcommand{\toolpong}[0]{\textsc{Pong}\xspace}
\newcommand{\tools}[0]{\textsc{PingPong's}\xspace}
\newcommand{\wansniffer}[0]{WAN sniffer\xspace}
\newcommand{\wifisniffer}[0]{Wi-Fi sniffer\xspace}
\newcommand{\event}[0]{event\xspace}
\newcommand{\events}[0]{events\xspace}
\newcommand{\timewindow}[0]{\textit{t}\xspace}

\newcommand{\PC}[0]{\ensuremath $P_c$}
\newcommand{\PCspace}[0]{\ensuremath $P_c$ }
\newcommand{\PCi}[0]{\ensuremath $P_{c_i}$}
\newcommand{\PCispace}[0]{\ensuremath $P_{c_i}$ }
\newcommand{\PCinext}[0]{\ensuremath $P_{c_{i+1}}$}
\newcommand{\PCinextspace}[0]{\ensuremath $P_{c_{i+1}}$ }
\newcommand{\PCinextnext}[0]{\ensuremath $P_{c_{i+2}}$}

\newcommand{\CPCi}[0]{\ensuremath $C-$\PCi}
\newcommand{\SPCi}[0]{\ensuremath $S-$\PCi}
\newcommand{\CPCinext}[0]{\ensuremath $C-$\PCinext}
\newcommand{\SPCinext}[0]{\ensuremath $S-$\PCinext}
\newcommand{\Pequals}[0]{\ensuremath $p=($}
\newcommand{\Pclose}[0]{\ensuremath $)$ }

\newcommand{\Pcomma}[0]{\ensuremath $, $ }
\newcommand{\Nil}[0]{\ensuremath $\mathrm{nil}$}

\newcommand{\tplinkplug}[0]{TP-Link plug\xspace}
\newcommand{\tplinktwooutletplug}[0]{TP-Link two-outlet plug\xspace}
\newcommand{\tplinkpowerstrip}[0]{TP-Link power strip\xspace}
\newcommand{\tplinkwhitebulb}[0]{TP-Link white light bulb\xspace}
\newcommand{\tplinkcamera}[0]{TP-Link camera\xspace}
\newcommand{\dlinkplug}[0]{D-Link plug\xspace}
\newcommand{\wemoplug}[0]{WeMo plug\xspace}
\newcommand{\wemoinsightplug}[0]{WeMo Insight plug\xspace}
\newcommand{\amazonplug}[0]{Amazon plug\xspace}
\newcommand{\dlinksiren}[0]{D-Link siren\xspace}
\newcommand{\tplinkbulb}[0]{TP-Link light bulb\xspace}
\newcommand{\lifxbulb}[0]{LiFX light bulb\xspace}
\newcommand{\ecobeethermostat}[0]{Ecobee thermostat\xspace}
\newcommand{\nestthermostat}[0]{Nest thermostat\xspace}
\newcommand{\arlocamera}[0]{Arlo camera\xspace}
\newcommand{\rachiosprinkler}[0]{Rachio sprinkler\xspace}
\newcommand{\blossomsprinkler}[0]{Blossom sprinkler\xspace}
\newcommand{\smartthingsplug}[0]{SmartThings plug\xspace}
\newcommand{\kwiksetdoorlock}[0]{Kwikset lock\xspace}
\newcommand{\huebulb}[0]{Hue light bulb\xspace}
\newcommand{\sengledbulb}[0]{Sengled light bulb\xspace}
\newcommand{\roombarobot}[0]{Roomba robot\xspace}
\newcommand{\ringalarm}[0]{Ring alarm\xspace}

\newcommand{\imcdataset}[0]{Mon(IoT)r\xspace}
\newcommand{\amazoncamera}[0]{Amazon camera\xspace}
\newcommand{\lefuncamera}[0]{Lefun camera\xspace}
\newcommand{\flexbulb}[0]{Flex light bulb\xspace}
\newcommand{\winkhub}[0]{Wink hub\xspace}
\newcommand{\zmododoorbell}[0]{ZModo doorbell\xspace}
\newcommand{\honeywellthermostat}[0]{Honeywell thermostat\xspace}

\newcommand{\microsevencamera}[0]{Microseven camera\xspace}
\newcommand{\samsungfridge}[0]{Samsung fridge\xspace}
\newcommand{\allurespeaker}[0]{Allure speaker\xspace}
\newcommand{\blinkhub}[0]{Blink hub\xspace}
\newcommand{\blinkcamera}[0]{Blink camera\xspace}
\newcommand{\insteonhub}[0]{Insteon hub\xspace}
\newcommand{\echodot}[0]{Amazon Echo Dot\xspace}
\newcommand{\echoplus}[0]{Amazon Echo Plus\xspace}
\newcommand{\echospot}[0]{Amazon Echo Spot\xspace}
\newcommand{\googlehome}[0]{Google Home\xspace}
\newcommand{\googlehomemini}[0]{Google Home Mini\xspace}

\newcommand{\firetv}[0]{Fire TV\xspace}
\newcommand{\lgtv}[0]{LG TV\xspace}
\newcommand{\rokutv}[0]{Roku TV\xspace}
\newcommand{\samsungtv}[0]{Samsung TV\xspace}

\newcommand{\localphone}[0]{Local-Phone\xspace}
\newcommand{\remotephone}[0]{Remote-Phone\xspace}
\newcommand{\ifttt}[0]{IFTTT\xspace}
\newcommand{\phonedevicecomm}[0]{Phone-Device\xspace}
\newcommand{\devicecloudcomm}[0]{Device-Cloud\xspace}
\newcommand{\phonecloudcomm}[0]{Phone-Cloud\xspace}

\theoremstyle{definition}
\newtheorem{definition}{Definition}[section]

\newcounter{numbers}		
\newcommand{\numbers}[0]{\stepcounter{numbers}\thenumbers.}

\usepackage[textsize=scriptsize]{todonotes} 
\begin{document}
\title{\tool: Packet-Level Signatures for Smart Home Device Events}

\author{
\IEEEauthorblockN{Rahmadi Trimananda, Janus Varmarken, Athina Markopoulou, Brian Demsky}
\IEEEauthorblockA{University of California, Irvine\\
\{rtrimana, jvarmark, athina, bdemsky\}@uci.edu}
}

\mycomment{
\IEEEoverridecommandlockouts
\makeatletter\def\@IEEEpubidpullup{6.5\baselineskip}\makeatother
\IEEEpubid{\parbox{\columnwidth}{
    Network and Distributed Systems Security (NDSS) Symposium 2020\\
    23-26 February 2020, San Diego, CA, USA\\
    ISBN 1-891562-61-4\\
    https://dx.doi.org/10.14722/ndss.2020.24097\\
    www.ndss-symposium.org
}
\hspace{\columnsep}\makebox[\columnwidth]{}}
}

\IEEEoverridecommandlockouts
\makeatletter\def\@IEEEpubidpullup{6.5\baselineskip}\makeatother
\IEEEpubid{\parbox{\columnwidth}{
		This is the technical report of the paper\\
		\textit{Packet-Level Signatures for Smart Home Devices} published at\\ 
		Network and Distributed Systems Security (NDSS) Symposium 2020\\
    23-26 February 2020, San Diego, CA, USA\\
}
\hspace{\columnsep}\makebox[\columnwidth]{}}

\maketitle

\begin{abstract}
Smart home devices are vulnerable to passive inference attacks 
based on network traffic, even in the presence of encryption.
In this paper, we present \tool, a tool that can automatically
extract packet-level signatures for  device events (\eg light bulb turning ON/OFF) from network traffic. 
We evaluated \tool on popular smart home devices ranging from smart plugs 
and thermostats to cameras, voice-activated devices, and smart TVs. 
 We  were able to: 
 (1) automatically extract  previously unknown signatures that consist of simple sequences of packet
lengths and directions;
 (2) use those signatures to detect the devices or specific events with an average recall of more than 97\%;  
 (3) show that the signatures are unique among hundreds of millions of packets of real world network traffic; 
 (4) show that our methodology is also applicable to publicly available datasets; and
 (5) demonstrate its robustness in different settings: 
 events triggered by local and remote smartphones, as well as by home-automation systems.

\end{abstract}

\section{Introduction}
\label{sect:introduction}

\athina{
Modern smart home devices are seeing widespread adoption.
They typically connect to the Internet via the home Wi-Fi router and can be controlled using a smartphone or voice assistant.
Although most modern smart home devices encrypt their network traffic, recent work has demonstrated that the smart home is susceptible to passive inference attacks \cite{princeton-spying-blinds,princeton-spying-castle,princeton-spying,princeton-stp,lopez-classifier,sivanathan-classifier,sivanathan-classifier-new,copos2016anybody,acar2018peek}. An eavesdropper may use characteristics of the network traffic generated by smart home  devices  to infer the device type
and activity, and eventually user behavior.
However, existing passive inference techniques have limitations.
Most can only identify the device type and whether there is device activity (an event), but not the exact type of event or command~\cite{princeton-spying-blinds,princeton-spying-castle,princeton-spying,princeton-stp,lopez-classifier,sivanathan-classifier,sivanathan-classifier-new}.
Others only apply to a limited number of devices from a specific vendor~\cite{copos2016anybody}, or 
need more information from other protocols (\eg Zigbee/Z-Wave)~\cite{acar2018peek,homonit} and the application source code~\cite{homonit}. 
Inference based on traffic volume analysis can be prevented by traffic shaping~\cite{acar2018peek,princeton-stp}.
Finally, many of these attacks assume that IP traffic is sniffed upstream from the home router, while the scenario where a local attacker sniffs encrypted Wi-Fi traffic has received less attention~\cite{ghiglieri2014,princeton-stp}.
}

In this paper, we experiment with  a  diverse range of smart home devices, namely 19 popular Wi-Fi and Zigbee devices (12 of which are the most popular smart home devices on Amazon) from 16 popular vendors, including smart plugs, light bulbs, thermostats, home security systems, \etc
During our  analysis of the network traffic that these devices generate, we observed that events on smart home devices typically result in communication between the device, the smartphone, and the cloud servers that contains pairs of packets with predictable lengths.
A packet pair typically consists of a \emph{request} packet from a device/phone (``\textsc{Ping}''), and a \emph{reply} packet back to the device/phone (``\textsc{Pong}'').
In most cases, the packet lengths are distinct for different device types and \events, thus, can be used to infer the device and 
the specific type of event that occurred. 
Building on this observation, we were able to identify new \emph{packet-level signatures} (or \emph{signatures} for short) that consist only of the lengths and directions of a few packets in the smart home device traffic.
In this paper, we show that these signatures: 
(1) can be extracted in an automated and systematic way without prior knowledge of the device's behavior;
(2) can be used to infer fine-grained information, \eg event types; 
(3) correspond to a variety of different \events (\eg ``toggle ON/OFF'' and ``Intensity''/``Color''); and 
(4) have a number of advantages compared to prior (\eg statistical, volume-based) approaches. 
More specifically, this paper makes the following contributions.

\myparagraph{New Packet-Level Signatures.}
\athina{We discover new  IoT device signatures that are simple and intuitive: they consist of short sequences of (\majorrevision{typically 2-6}) packets of specific lengths,} exchanged between the device, the smartphone, and the cloud. 
The signatures are effective:
\squishcount
\item
They detect event occurrences with an average recall of more than 97\%, surpassing the state-of-the-art 
techniques (see Sections~\ref{sect:related-work} and~\ref{sect:detection-phase}). 
\item
\majorrevision{
They are unique: we observe a low false positive rate (FPR), namely  1 false positive per 40 million packets in network traces with hundreds of millions of packets (see Section~\ref{sect:negative-control}).} 
\item
\majorrevision{
They characterize a wide range of devices: 
(i) we extract signatures for 18 out of the 19 devices we experimented with, including the most popular home security 
devices such as the Ring Alarm Home Security System and Arlo Q Camera (see Section~\ref{sect:training-phase});
(ii) we extract signatures for 21 additional devices from a public dataset~\cite{ren2019information}, including
more complex devices, \eg voice-command devices, smart TVs, and even a fridge (see Section~\ref{sect:public-dataset}).
}
\item
\majorrevision{
They are robust across a diverse range of settings: 
(i)~we extract signatures both from testbed experiments and publicly available datasets; and
(ii)~we trigger events in different ways, \ie using both a local and a remote smartphone, and using a home automation system.  
}
\item
They can be extracted from both unencrypted and encrypted traffic. 
\item
They allow quick detection of \events as they rely only on a few packet lengths and directions, and do not require any statistical computation. 
\countend

\myparagraph{Automated Extraction of Packet-Level Signatures.}
We present \tool, a methodology and software tool that: 
(1) automates the extraction of packet-level signatures without prior knowledge about the device, and 
(2) detects signatures in network traces and real network traffic.
For signature extraction, \tool first generates training data by repeatedly triggering the event, for which a signature is desired, while capturing network traffic.
Next, \tool extracts request-reply packet pairs per flow (``\textsc{Ping}-\textsc{Pong}''), clusters these pairs, and post-processes them to concatenate pairs into longer sequences where possible. 
Finally, \tool selects sequences with frequencies close to the number of triggered events as the final signatures.
The signature detection part of \tool{} leverages the simplicity of packet-level signatures and is implemented using simple state machines.
\tool's implementation and datasets are made available at~\cite{pingpong-software}.

The remainder of this paper is structured as follows.  
Section~\ref{sect:related-work} outlines related work and puts \tool{} in perspective.  
Section~\ref{sect:problem-setup} presents the threat model (including two distinct adversaries: a \wansniffer and a \wifisniffer), our experimental setup, and an illustrative example of packet-level signatures in smart plugs.
Section~\ref{sect:system-pipeline} presents the design of the \tool system, including extraction and detection of signatures.
Section~\ref{sect:evaluation} presents the evaluation of \tool, using our own testbed experiments, as well as several external---publicly available---datasets.
Section~\ref{sect:defenses} presents an in-depth discussion on possible defenses against packet-level signatures.
Section~\ref{sect:conclusion} concludes and outlines directions for future work.

\section{Related Work}
\label{sect:related-work}

Table~\ref{tab:summary-contributions} summarizes the properties of \tool and compares it to the 
other IoT traffic analysis approaches.

\myparagraph{Network Signatures for IoT devices.}
A growing body of work uses network traffic (metadata) analysis to characterize the type and activity of IoT devices.
A series of papers by Apthorpe \etal{}~\cite{princeton-stp,princeton-spying-blinds,princeton-spying-castle,princeton-spying} use traffic volume/shape-based signatures to infer IoT device activity, but cannot always determine the exact type of the \event{}.
Furthermore, the signatures corresponding to  different traffic shapes are intuitive, but not automatically extracted.
The authors propose stochastic traffic padding (STP) to mitigate volume-based inference attacks.

\majorrevision{
	HomeSnitch~\cite{homesnitch} by OConnor \etal{} identifies IoT activity using a key observation that is similar to ours, \ie the client (the IoT device) and the server take turns in a request-reply communication style. 
	HomeSnitch and \tool{} both exclude IP addresses, port numbers, and DNS information from their \event{} inference methodologies, but differ in terms of the granularity of the features they use: HomeSnitch uses statistics derived from the entire client-server dialog, whereas \tool{} considers the direction and length of each individual packet.
	Interestingly, the most important feature used in HomeSnitch is the average number of bytes sent from the IoT device to the server per turn.
	This result aligns with the main observation of this paper, \ie{} packet lengths of individual requests (and replies) uniquely identify device \events{}.
}



\majorrevision{
	A recent paper by Ren \etal{}~\cite{ren2019information} presents a large-scale measurement study of IoT devices and reveals how these devices operate differently in the US and the UK with respect to Internet endpoints contacted, exposure of private information, \etc{} We use that dataset to evaluate our methodology in Section~\ref{sect:public-dataset}. The paper also presents a classifier that can infer \event{} types spanning many device categories; this, however, is not the focus of the paper. 
	Other well-known measurement studies and publicly available IoT network traffic datasets include YourThings~\cite{sok-alrawi,yourthings} and~\cite{sivanathan-classifier}, which we use in our evaluation in Section~\ref{sect:negative-control}.
}


\begin{table}[t!]
  \centering
  \begin{center}
  { \footnotesize
  \begin{tabular}{| p{7mm} | C{7mm} | C{6mm} | c | c | c | c | C{7mm} |}
  	\hline
		& \multicolumn{7}{ c |}{Approaches for IoT Network Traffic Signatures}\\
	\cline{2-8}
	  & Vol. & Nest & \multicolumn{3}{ c |}{\majorrevision{Machine  Learning}} & ZigBee/& \cellcolor{green} \toolping\\
  \cline{4-6}
		& +DNS & device & \majorrevision{\cite{homesnitch}} & \majorrevision{\cite{acar2018peek}} & \cite{sivanathan-classifier-new} & Z-Wave & \cellcolor{green} \toolpong\\
		& based & \cite{copos2016anybody} & & & \cite{sivanathan-classifier} & device & \cellcolor{green} \\
		& \cite{princeton-spying-castle,princeton-spying,princeton-spying-blinds,princeton-stp} & & & & & \cite{homonit} & \cellcolor{green} \\
    \hline
    	\multicolumn{8}{| c |}{(1) Signature can detect}\\
    \hline
    	Device & \checkmark & \checkmark & \majorrevision{\checkmark} &  \checkmark & \checkmark & \checkmark & \cellcolor{green} \checkmark\\
		type & & & & & & & \cellcolor{green} \\
    \hline
    	Event & $\times$ & \checkmark & \majorrevision{\checkmark} & \checkmark & $\times$ & \checkmark & \cellcolor{green} \checkmark\\
    	type & & & & & & & \cellcolor{green} \\
    \hline
    	\multicolumn{8}{| c |}{(2) Applicability to devices}\\
    \hline
    	$>15$ & $\times$ & $\times$ & \majorrevision{\checkmark} &  \checkmark & \checkmark & $\times$ & \cellcolor{green} \checkmark \\
    	Models & & & & & & & \cellcolor{green}\\
    \hline
    	\multicolumn{8}{| c |}{(3) Observation points/threat models}\\
  	\hline
    	LAN & $\times$ & \checkmark & \majorrevision{\checkmark} &  \checkmark & \checkmark & N/A & \cellcolor{green} \checkmark\\
  	\hline
    	WAN & \checkmark & $\times$ & \majorrevision{$\times$} & $\times$ & $\times$ & N/A & \cellcolor{green} \checkmark\\
  	\hline
    	Wi-Fi & \checkmark & $\times$ & \majorrevision{$\times$} & \checkmark & $\times$ & N/A & \cellcolor{green} \checkmark\\
    \hline
    	\multicolumn{8}{| c |}{(4) Signature characteristics}\\
    \hline
    	Feat. & Traffic & TCP & \majorrevision{13}, & \majorrevision{(795)} & 12 & Packet & \cellcolor{green} Packet \\
    	& vol.,  & conn. & \majorrevision{ADU} & \majorrevision{197} & & length & \cellcolor{green} length \\
		& DNS & size,   &                              &  & & \& dir.& \cellcolor{green} \& dir.\\
		& & proto. & & & &  & \cellcolor{green} \\
	\hline
    	Inter- & \checkmark & $\times$ & \majorrevision{\checkmark} & $\times$ & $\times$ & \checkmark & \cellcolor{green} \checkmark \\
    	pretable & & & & & & & \cellcolor{green} \\
	\hline
    	Auto. & $\times$ & $\times$ & \majorrevision{\checkmark} & \checkmark & \checkmark & \checkmark & \cellcolor{green} \checkmark \\
    	Extract. & & & & & & & \cellcolor{green} \\
    \hline
    	\multicolumn{8}{| c |}{(5) Resilient against defenses}\\
    \hline
    	VPN & $\times$ & $\times$ & \majorrevision{$\times$} & $\times$ & $\times$ & N/A & \cellcolor{green} \checkmark\\
    \hline
    	Traffic & $\times$ & $\times$ & \majorrevision{$\times$} & $\times$ & $\times$ & N/A & \cellcolor{green} \checkmark\\
    	shaping & & & & & & & \cellcolor{green} \\
    	
    \hline
  \end{tabular}
  }
  \end{center}
  \caption{ \tool's properties vs. alternative approaches (\checkmark = Yes; $\times$ = No).
  	\label{tab:summary-contributions}}
  	\vspace{-2.5em}
\end{table}

Other papers consider specific types of devices or protocols. 
Copos \etal{}~\cite{copos2016anybody} analyze network traffic of the Nest Thermostat and Nest Protect (only) and show that the thermostat's transitions between the \emph{Home} and \emph{Auto Away} modes can be inferred. 
Other work~\cite{acar2018peek,homonit} focus on Zigbee/Z-Wave devices and leverage specialized Zigbee/Z-Wave sniffers. 

Most \event inference techniques rely on machine learning~\cite{lopez-classifier,sivanathan-classifier,sivanathan-classifier-new} or statistical analysis of traffic time series~\cite{copos2016anybody,acar2018peek,homesnitch,ren2019information}. 
Limitations of these approaches include: the inability to differentiate \event{} \emph{types}~\cite{lopez-classifier,sivanathan-classifier,sivanathan-classifier-new} (\eg{} distinguishing ON from OFF), and lack of resistance to traffic shaping techniques~\cite{copos2016anybody,acar2018peek,homesnitch,ren2019information} such as~\cite{princeton-stp}.
On the other hand, our work identifies simple packet exchange(s) between the device/smartphone and the cloud that uniquely identify event types.
At the same time, \tool{}'s classification performance (recall of more than 97\%) is better than most statistical  approaches: \cite{acar2018peek} reported 90\% accuracy,~\cite{copos2016anybody} reported 88\% and 67\% accuracy, and~\cite{ren2019information} reported some F1 scores as low as 0.75.
Unsupervised learning techniques may be hard to interpret, especially for large feature sets (\eg 197 features in~\cite{acar2018peek}).
\tool also uses clustering to identify reoccurring packet pairs, but provides an intuitive interpretation of those pairs: they correspond to a request and the subsequent reply.

  \myparagraph{Network Traffic Analysis beyond IoT.}  
  There is a large body of work in the network measurement community that uses traffic analysis to classify applications and identify anomalies~\cite{nguyen-survey-ml,karagiannis-blinc,jin-unveiling}, attacks~\cite{ml-ddos}, or malware~\cite{anderson-identifying-encrypted-malware,perdisci-http-malware}.
  There has also been a significant amount of work on fingerprinting techniques in the presence of encryption for web
  browsing~\cite{bissiaspets,herrmann2009website,liberatore2006inferring,chen2010side,
  luchang,panchenko2011website,dyer2012peek,cai2012,wang2014effective,panchenko2016website,hayes2016},
  and variable bit-rate encodings for
  communication~\cite{wrightusenix,wright2010} and
  movies~\cite{saponas}.  
  For these examples, the underlying protocols are well understood, while \tool 
  can work with (and is agnostic to) any arbitrary, even proprietary, application-layer protocol. 
  
  \majorrevision{
  \myparagraph{Defenses.}
  	Related to profiling and fingerprinting is also the body of work on defenses that obfuscate traffic signatures. 
  	Examples include~\cite{panchenko2011website, luo2011} that use packet padding and traffic injection techniques to prevent website fingerprinting.
	In Table~\ref{tab:summary-contributions}, we mention two general defense approaches: (1)~\emph{traffic shaping} that refers broadly to changing the shape of traffic over time; and (2)~\emph{VPN} that brings multiple benefits such as encryption (that our signatures survive), and multiplexing of several flows. We partly evaluate these defenses (see \appdx{Appendix~\ref{sect:defenses-packet-injection-cr}}). A VPN also provides a natural place to implement additional defenses (\eg packet padding, which is discussed in Section~\ref{sect:defenses}).
}

\begin{figure}[t!]
    \centering
	\includegraphics[width=1.\linewidth]{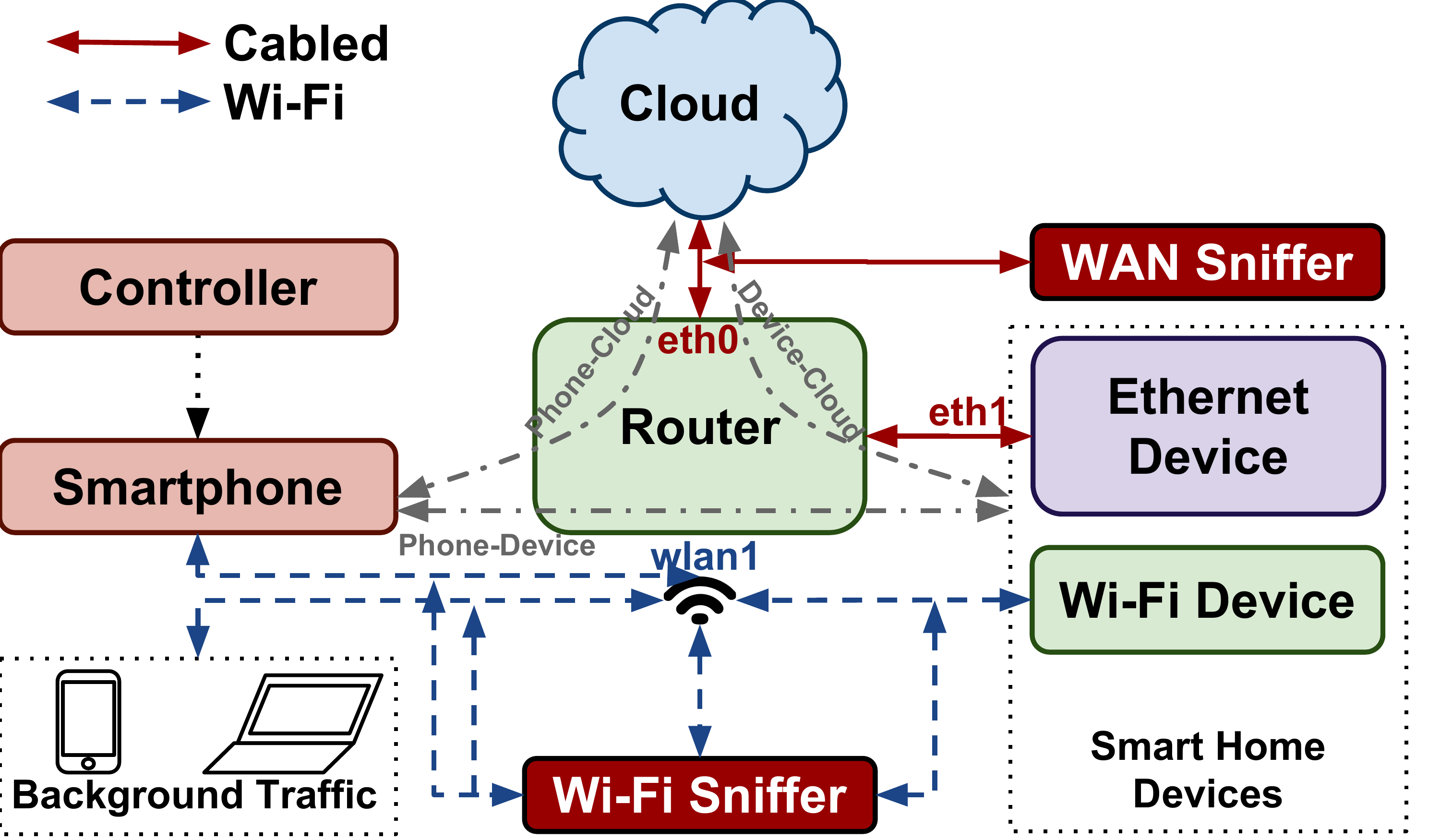}
	\small
    \caption{Our experimental setup for studying smart home devices.
    	``Wi-Fi Device'' is any smart home device connected to the router via Wi-Fi 
    	(\eg Amazon and WeMo plugs).
    	``Ethernet Device'' is any smart home device connected to the router via Ethernet 
    	(\eg SmartThings hub that relays the communication of Zigbee devices).
    	Smart home device \events  may result in communication between
		\phonecloudcomm, \devicecloudcomm,
    	 or \phonedevicecomm.
	There may also be background traffic from additional computing devices in the home.
	 \label{fig:combined-setup}}
	 \vspace{-1em}
	\end{figure}
	
\begin{table}[htb!]
	\centering
	\begin{center}
		{ \footnotesize
			\begin{tabular}{| r | p{25mm} | p{43mm} |}
				\hline
				\multicolumn{1}{| r |}{\textbf{No.}} & \textbf{Device Name} & \multicolumn{1}{ p{43mm} |}{\textbf{Model Details}} \\
				\hline
				\rowcolor{green} \multicolumn{1}{| r |}{\numbers} & \amazonplug & \multicolumn{1}{ p{43mm} |}{Amazon Smart Plug} \\
				\hline
				\rowcolor{green} \multicolumn{1}{| r |}{\numbers} & \wemoplug & \multicolumn{1}{ p{43mm} |}{Belkin WeMo Switch} \\
				\hline
				\rowcolor{green} \multicolumn{1}{| r |}{\numbers} & \wemoinsightplug & \multicolumn{1}{ p{43mm} |}{Belkin WeMo Insight Switch} \\
				\hline
				\rowcolor{green} \multicolumn{1}{| r |}{\numbers} & \sengledbulb & \multicolumn{1}{ p{43mm} |}{Sengled Element Classic} \\
				\hline
				\rowcolor{green} \multicolumn{1}{| r |}{\numbers} & \huebulb & \multicolumn{1}{ p{43mm} |}{Philips Hue white} \\
				\hline
				\rowcolor{green} \multicolumn{1}{| r |}{\numbers} & \lifxbulb & \multicolumn{1}{ p{43mm} |}{LiFX A19} \\
				\hline
				\rowcolor{green} \multicolumn{1}{| r |}{\numbers} & \nestthermostat & \multicolumn{1}{ p{43mm} |}{Nest T3007ES} \\
				\hline
				\rowcolor{green} \multicolumn{1}{| r |}{\numbers} & \ecobeethermostat & \multicolumn{1}{ p{43mm} |}{Ecobee3} \\
				\hline
				\rowcolor{green} \multicolumn{1}{| r |}{\numbers} & \rachiosprinkler & \multicolumn{1}{ p{43mm} |}{Rachio Smart Sprinkler Controller Generation 2} \\
				\hline
				\rowcolor{green} \multicolumn{1}{| r |}{\numbers} & \arlocamera & \multicolumn{1}{ p{43mm} |}{Arlo Q} \\
				\hline
				\rowcolor{green} \multicolumn{1}{| r |}{\numbers} & \roombarobot & \multicolumn{1}{ p{43mm} |}{iRobot Roomba 690}\\
				\hline
				\rowcolor{green} \multicolumn{1}{| r |}{\numbers} & \ringalarm & \multicolumn{1}{ p{43mm} |}{Ring Alarm Home Security System} \\
				\hline
				\multicolumn{1}{| r |}{\numbers} & \tplinkplug & \multicolumn{1}{ p{43mm} |}{TP-Link HS-110} \\
				\hline
				\multicolumn{1}{| r |}{\numbers} & \dlinkplug & \multicolumn{1}{ p{43mm} |}{D-Link DSP-W215} \\
				\hline
				\multicolumn{1}{| r |}{\numbers} & \dlinksiren & \multicolumn{1}{ p{43mm} |}{D-Link DCH-S220} \\
				\hline
				\multicolumn{1}{| r |}{\numbers} & \tplinkbulb & \multicolumn{1}{ p{43mm} |}{TP-Link LB-130} \\
				\hline
				\multicolumn{1}{| r |}{\numbers} & \smartthingsplug & \multicolumn{1}{ p{43mm} |}{Samsung SmartThings Outlet (2016 model)} \\
				\hline
				\multicolumn{1}{| r |}{\numbers} & \kwiksetdoorlock & \multicolumn{1}{ p{43mm} |}{Kwikset SmartCode 910} \\
				\hline
				\multicolumn{1}{| r |}{\numbers} & \blossomsprinkler & \multicolumn{1}{ p{43mm} |}{Blossom 7 Smart Watering Controller} \\
				\hline
			\end{tabular}
		}
	\end{center}
	\caption{The set of smart home devices considered in this paper. Devices highlighted in green are among the most popular on Amazon. 
		\label{tab:smarthome-devices}}
	\vspace{-2em}
\end{table}

\begin{table*}[t!]
  \centering
  \begin{center}
  { 
  \small
  \begin{tabular}{| c | c | c |}
    \hline
        \multicolumn{1}{| c |}{\textbf{TP-Link}} & \textbf{D-Link} & \multicolumn{1}{ c |}{\textbf{SmartThings}}\\
    \hline
        \multicolumn{1}{| c |}{\includegraphics[width=0.3\linewidth, valign=T]{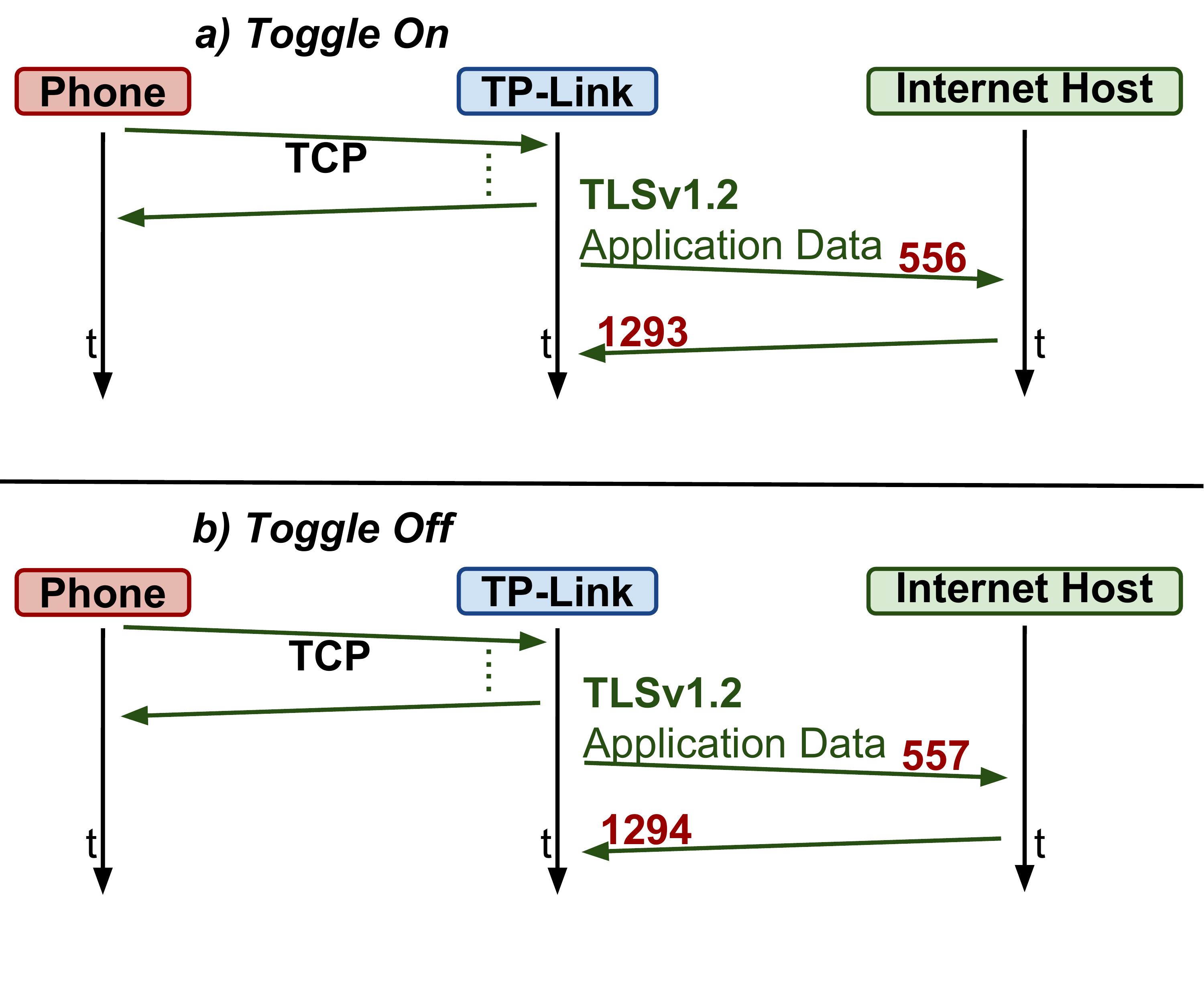}}
        & 
        \includegraphics[width=0.24\linewidth, valign=T]{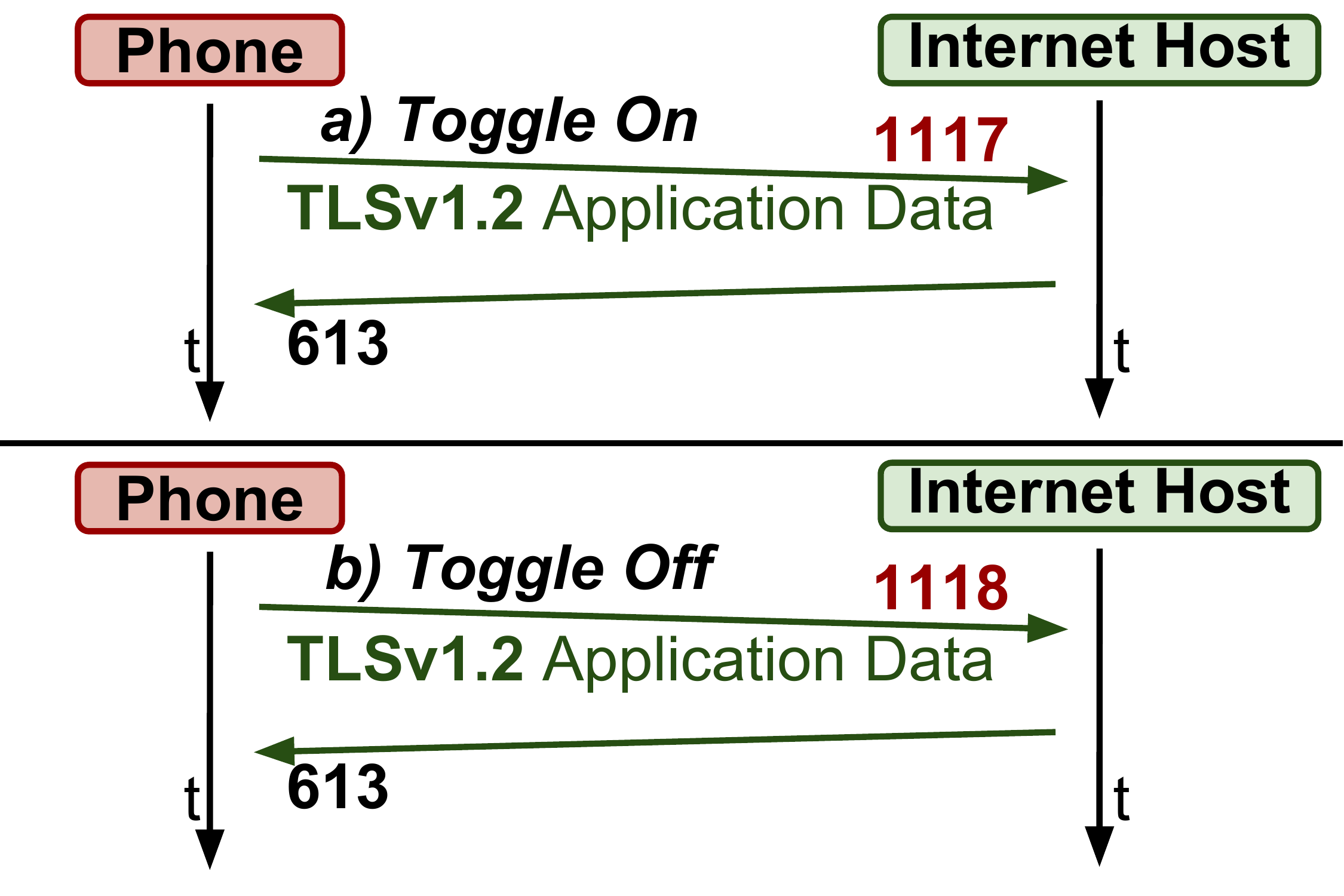}
        &
        \multicolumn{1}{ c |}{\includegraphics[width=0.38\linewidth, valign=T]{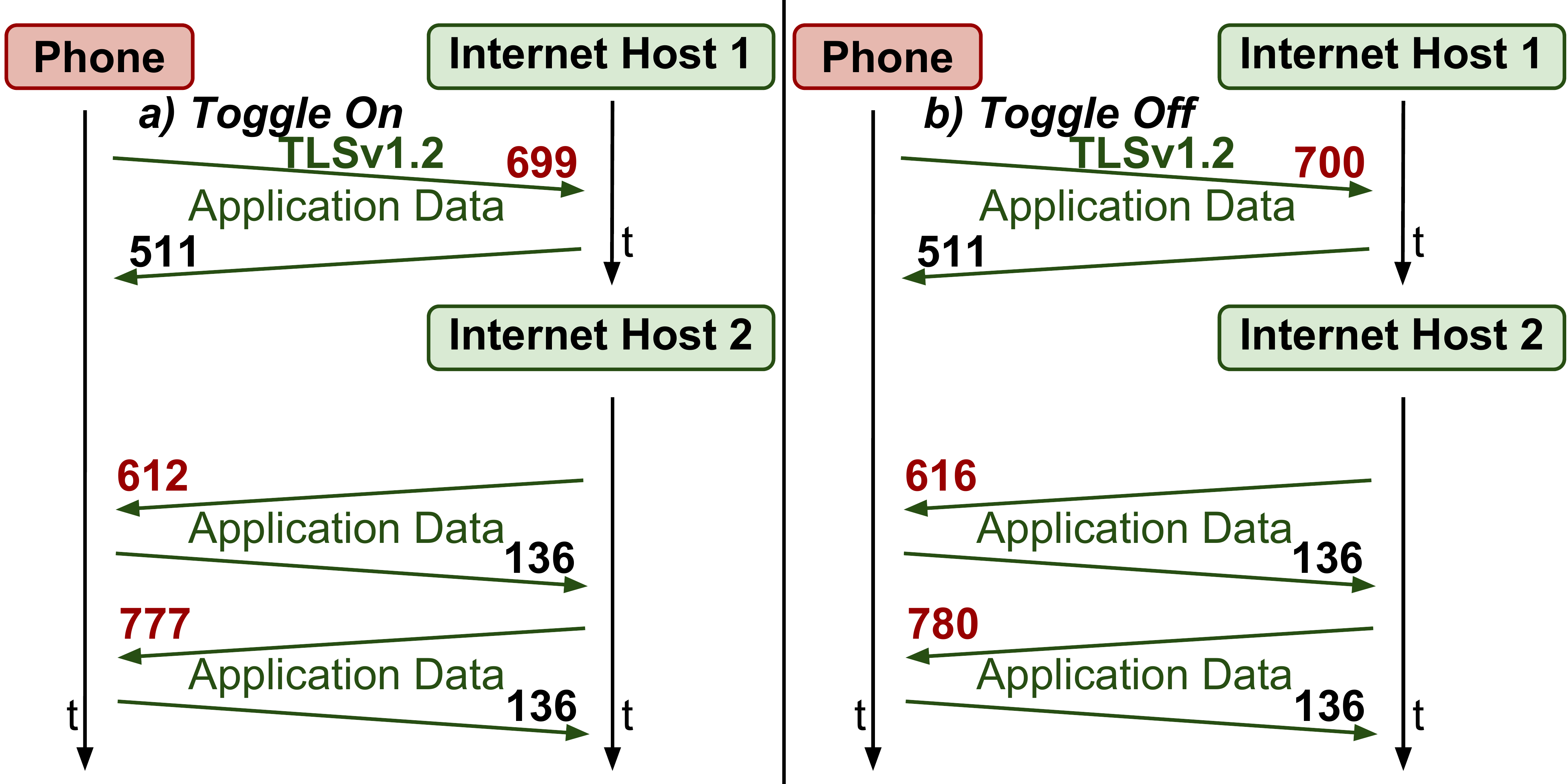}}\\
    \hline
  \end{tabular}
  }
\end{center}
\caption{Packet-level signatures of TP-Link, D-Link, and SmartThings smart plugs 
observable by the WAN sniffer. The numbers represent packet lengths, with red indicating that the length is different for ON vs. OFF, and the arrows represent packet directions.\label{tab:packet-level-signatures}}
\vspace{-2em}
\end{table*}

\section{Problem Setup} \label{sect:problem-setup}

In this section, we first present our threat model.
Then, we present the smart home environment and the passive inference attacks we consider. We also discuss  
a key insight  we obtained from manually analyzing network traffic from the simplest devices---smart plugs.  The packet sequences we observed  in smart plugs
inspired the
\tool methodology for automatically extracting signatures.

\subsection{Threat Model}
\label{sect:threat-model}

In this paper, we are concerned with the network traffic of smart home devices
leaking private information about smart home devices and users.  Although
most smart home devices encrypt their communication, information can be leaked by traffic metadata such as the
lengths and directions of these encrypted packets.

We consider two different types of adversaries: a \emph{\wansniffer}
and a \emph{\wifisniffer}.  The adversaries differ in terms of the
vantage point where traffic is inspected and, thus, what information is
available to the adversary.
%
The \wansniffer monitors network traffic in the communication between the home router and 
the ISP network (or beyond)~\cite{princeton-spying-blinds,princeton-spying-castle,princeton-spying,princeton-stp}.
This adversary can inspect the IP headers of all packets, but does not know the device MAC addresses to identify which device has sent the
traffic.  We assume a standard home network that uses NAT: all
traffic from the home is multiplexed onto the router's IP address.
Examples of such adversaries include intelligence agencies and ISPs.
The \wifisniffer monitors encrypted IEEE 802.11
traffic, and has not been as widely studied~\cite{ghiglieri2014,
  princeton-stp}.  We assume that the \wifisniffer does not
know the WPA2 key, and thus only has access to the information sent in
clear text---the MAC addresses, packet lengths, and timing information.  As packets are encrypted, the
\wifisniffer does not have access to network and transport layer
information.

For both adversaries, we assume that the adversary 
knows the type of the smart home device that they wish to target and passively monitor.  
Thus, they can train the system on another device of the same type offline, 
extract the signature of the device, and perform the detection of the signature on the traffic coming
from the smart home they target. We assume that the
devices encrypt their communication and thus neither adversary has
access to the clear-text communication.

\subsection{Smart Home Environment and Experimental Testbed}
\label{sect:experimental-testbed}
\myparagraph{Experimental Testbed.}
Figure~\ref{fig:combined-setup} depicts our experimental setup, which resembles a typical smart home 
environment.
\mycomment{
\footnote{``Wi-Fi Device'' is any smart home device connected to the router via Wi-Fi 
    	 (\eg Amazon plug).
    	 ``Ethernet Device'' is any smart home device connected to the router via Ethernet 
    	 (\eg SmartThings hub that relays the communication of Zigbee devices).
    	 Smart home device commands  may result in communication between \emph{\phonecloudcomm}, \emph{\devicecloudcomm},
    	 or \emph{\phonedevicecomm}.}
		There may also be \emph{background traffic} due to additional computing devices at home.}
We experiment with 19 widely-used smart home devices from 
16 different vendors (see Table~\ref{tab:smarthome-devices}).
\majorrevision{
We attempted to select
a set of devices with a wide range of functionality---from plugs to cameras.} 
They are also widely used: these devices are popular and they come from well-known vendors.
The first 12 (highlighted in green) are the most popular on 
Amazon~\cite{topdevices}: (1)~each received the most reviews for its respective device type and
(2) each had at least a 3.5-star rating---they are both popular and of high quality
(\eg the Nest T3007ES and Ecobee3 thermostats are the two most-reviewed with 4-star rating for thermostats).
Some devices are connected to the router via Wi-Fi (\eg the Amazon plug) and others through Ethernet. 
The latter includes the SmartThings, Sengled, and Hue hubs that relay communication to/from
Zigbee/Z-Wave devices: the SmartThings plug, Kwikset doorlock, Sengled light bulb, and Hue light bulb.


\jvedit{
Each smart home device in Figure~\ref{fig:combined-setup} is controlled from the smartphone using its vendor's official Android application.
In Figure~\ref{fig:combined-setup}, the smartphone is connected to a local network, which the devices are also connected to. 
When the smartphone is connected to a remote network, only the \devicecloudcomm communication is observable in the local network---the smartphone controls a device by communicating with its vendor-specific cloud, and the cloud relays the command to the device.
The controller represents the agent that operates the smartphone to control the smart home device  of interest. This may be done manually by a human (as in Section~\ref{sect:network-behavior}) or through software (as in Section~\ref{sect:system-pipeline}). Additionally, there are other computing devices (\eg laptops, tablets, phones) in the house that also generate network traffic, which we refer to as ``Background Traffic''.
The router runs OpenWrt/LEDE~\cite{openwrt}, a Linux-based OS for network devices, and serves as our vantage point for collecting traffic for experiments. 
We run \code{tcpdump} on the router's WAN interface (\code{eth0}) and local interfaces (\code{wlan1} and \code{eth1}) to capture Internet traffic as well as local traffic for all Wi-Fi and Ethernet devices.
We use the testbed to generate training data for each device, from which we in turn extract signatures (Section~\ref{sect:training-phase}). 
In Section \ref{sect:detection-phase}, the same testbed is used for testing, \ie to detect the presence of 
the extracted signatures in traffic generated by all the devices as well as by other computing devices (background traffic). 
}


\mycomment{
\begin{figure}[t!]
  \centering
	\includegraphics[width=1.\linewidth]{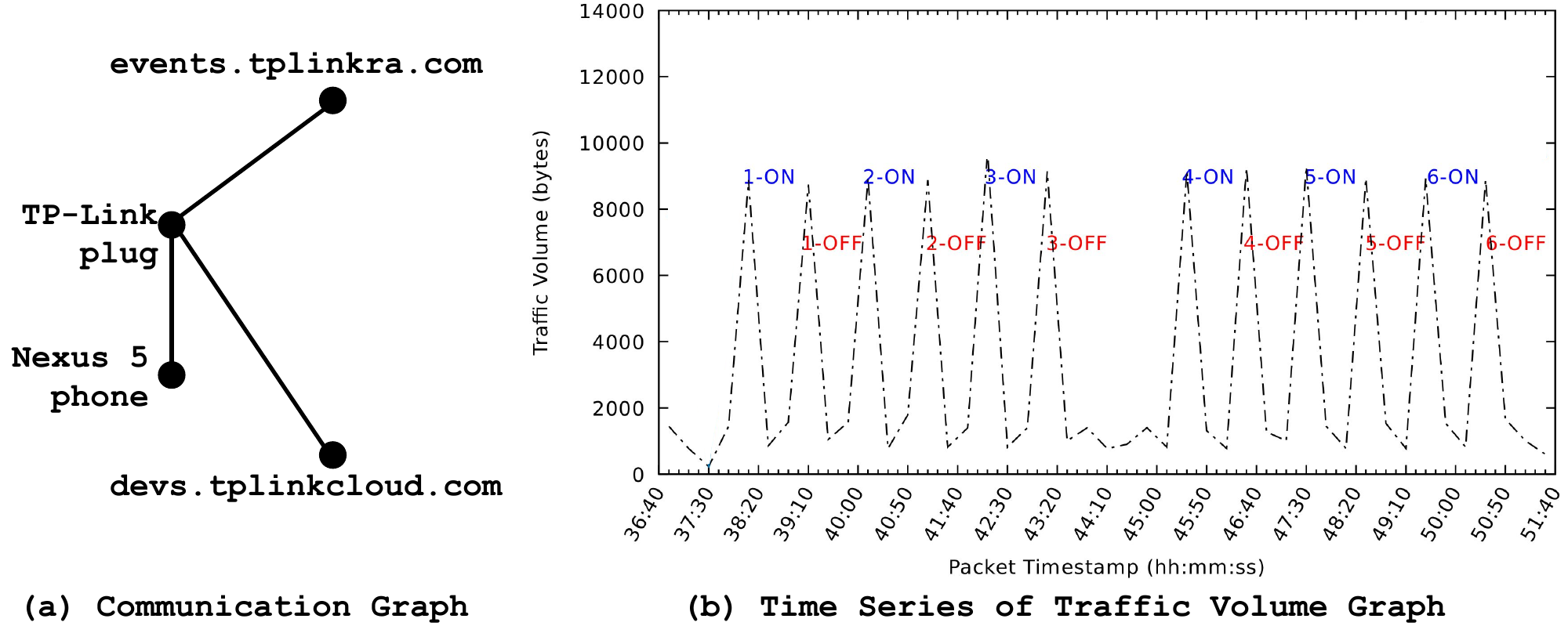}
\caption{\footnotesize Communication and time series of traffic volume graphs for the TP-Link smart plug,  during the experiment with 3 ON and 3 OFF commands.
\label{fig:plugs-signature-tplink}}
\vspace{-10pt}
\end{figure}
}


\rahmedit{
\myparagraph{Communication.}
Smart home device \events  may result in communication between three different pairs of devices, as depicted in Figure~\ref{fig:combined-setup}: (1)~the smartphone and the smart home device (\emph{\phonedevicecomm}); (2)~the smart home device and an Internet host (\emph{\devicecloudcomm}), and (3)~the smartphone and an Internet host (\emph{\phonecloudcomm}). The idea behind a passive inference attack is that network traffic on any of these three communication paths 
may contain unique traffic signatures that can be exploited to infer the occurrence of \events.
}

\subsection{Motivating Case: Smart Plugs}
\label{sect:network-behavior}
As an illustrative example, let us discuss our manual analysis of 3 smart plugs: the \tplinkplug, the \dlinkplug, and the \smartthingsplug. 
Data for the manual analysis was collected using the setup in Figure~\ref{fig:combined-setup}. 
For each device, we toggled it ON, waited for approximately one minute, and then toggled it OFF.
This procedure was repeated for a total of 3 ON and 3 OFF events, separated by one minute in between.
Timestamps were manually noted for each \event.
The PCAP files logged at the router were analyzed using a combination of scripts and manual inspection in Wireshark.

\myparagraph{New Observation: Packet Pairs.}
We  identified the traffic flows that occurred immediately after each \event and observed that certain pairs of packets with specific lengths and directions followed each ON/OFF \event: the same pairs consistently showed up for all \events of the same type (\eg ON), but were slightly different across event types (ON vs. OFF).
The pairs were comprised of a \emph{request} packet in one direction, and a \emph{reply} packet in the opposite direction.
Intuitively, this makes sense: if the smart home device changes state, this information needs to be sent to (request), and acknowledged by (reply), the cloud server to enable devices that are not connected to the home network to query the smart home device's current state. 
These exchanges resemble the ball that moves back and forth between players in a game of pingpong, which inspired the name for our software tool.

Table~\ref{tab:packet-level-signatures} illustrates the observed packet exchanges.
For the \tplinkplug, we observed an exchange of 2 TLS Application Data packets between the plug and an Internet host where the packet lengths were 556 and 1293 when the plug was toggled ON, but 557 and 1294 for OFF.
We did not observe any pattern in the \dlinkplug's own communication.
However, for ON \events, the controlling smartphone would always send a request packet of length 1117 to an Internet host and receive a reply packet of length 613.
For OFF, these packets were of lengths 1118 and 613, respectively.
Similarly for the \smartthingsplug, we found consistently occurring packet pairs in the smartphone's communication with two different Internet hosts where the lengths of the request packets were different for ON and OFF \events. 
Thus, this request-reply pattern can occur in the communication of any of the three pairs: \phonedevicecomm, \devicecloudcomm, or 
\phonecloudcomm (see Figure~\ref{fig:combined-setup}).

\myparagraph{Key Insight.}
This preliminary analysis indicates that each type of \event is uniquely identified by the exchange of pairs (or longer sequences) of packets of specific 
lengths. 
To the best of our knowledge, this type of network signature has not been observed before, and we refer to it as a \emph{packet-level signature}.

\section{\tool Design}
\label{sect:system-pipeline}

The key insight obtained from our manual analysis in Section~\ref{sect:network-behavior} was that unique sequences of packet lengths (for packet pairs or longer packet sequences) typically follow simple events (\eg ON vs. OFF) on smart plugs, and can potentially be exploited as signatures to infer these events.  
This observation motivated us to investigate whether: (1)  more  smart home devices, and potentially the smartphones that control them as well, exhibit their own unique packet-level sequences following an event, (2)  these signatures can be learned and automatically extracted, and (3)  they are sufficiently unique to accurately detect \events.  In this section, we present the design of \tool---a system that addresses the above questions with a resounding YES! 

\begin{figure}[t!]
    \centering
	\includegraphics[width=1\linewidth]{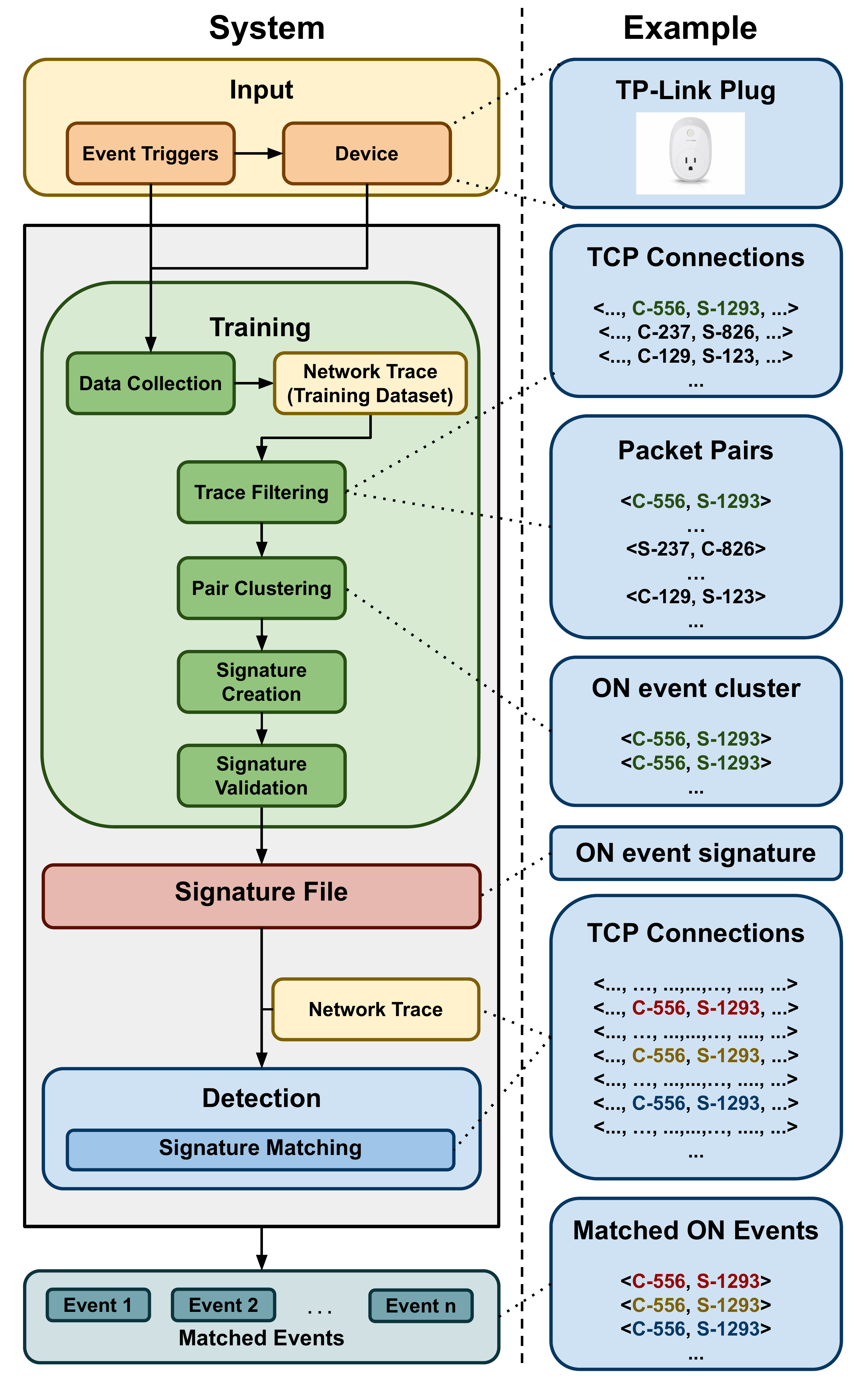}
    \caption{Left: \tool Overview. Right: \tplinkplug is used as a running example throughout this section.\label{fig:inspecto}}
\end{figure}

\rahmedit{
\tool automates the collection of  training data, extraction of packet-level signatures, and 
detection of the occurrence of a signature in a network trace.
\tool has two components: (1) training (Section~\ref{sect:system-training}), and
(2) detection (Section~\ref{sect:system-detection}).}
Figure~\ref{fig:inspecto} shows the building blocks and flow of \tool
on the left-hand side, and the TP-Link plug as an example on the right-hand side.
We use the latter as a running example throughout this section. 

\begin{figure*}[t!]
    \centering
	\includegraphics[width=0.95\linewidth]{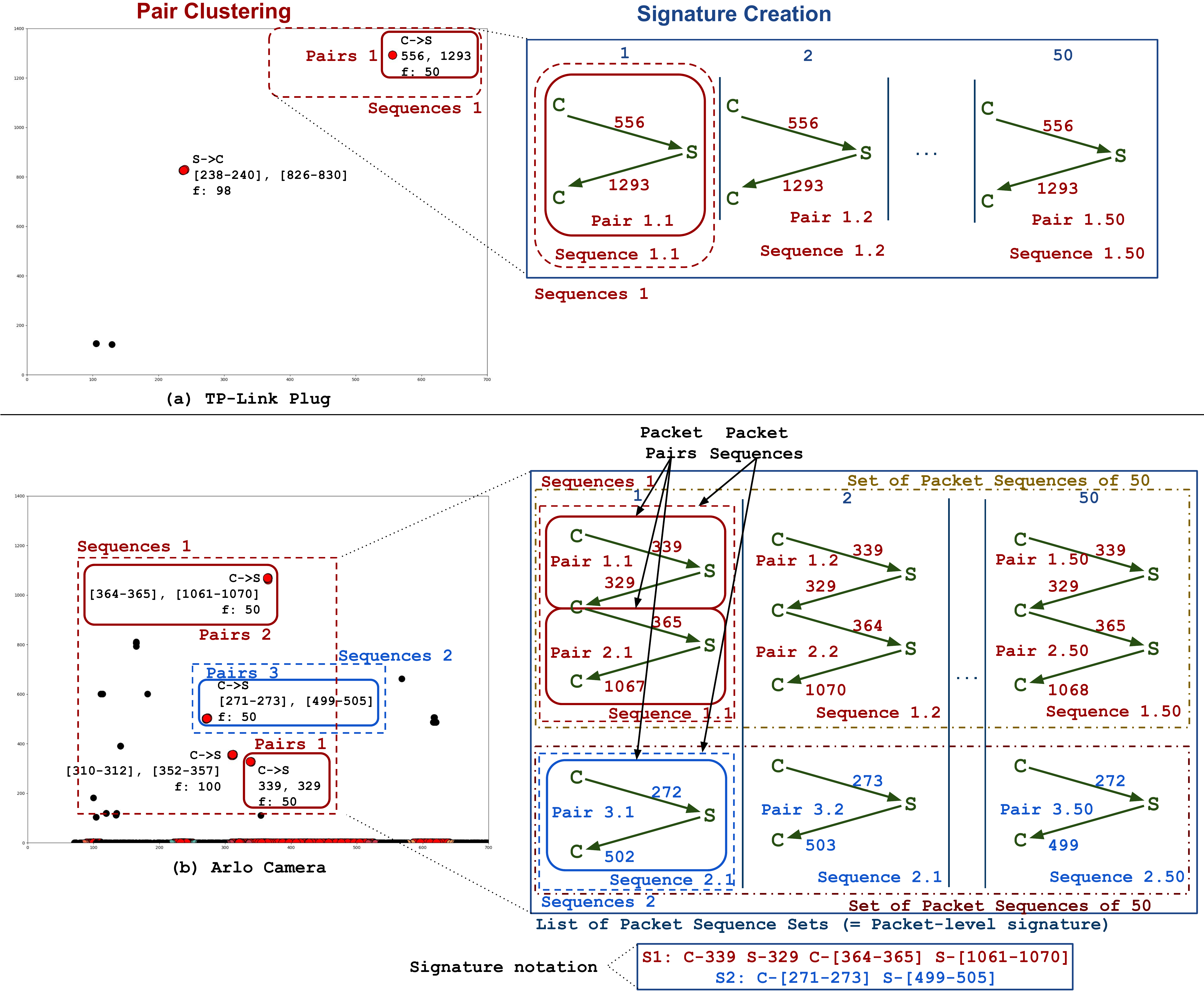}
    \caption{Pair clustering and signature creation for 2 extreme cases---\tplinkplug
    	has the simplest signature with only 1 pair (see our initial findings in
    	Table~\ref{tab:packet-level-signatures}). 
    	The \arlocamera has a more complex signature with 1 sequence of 2 pairs and 1 sequence of 1 pair.
    	The left subfigure, in every row, depicts the packet lengths in one packet pair ($P_{c_1}$, $P_{c_2}$).
    	Notation: \code{C->S} means a pair where the first packet's direction is \code{C}lient-to-\code{S}erver
    	and the second packet's direction is server-to-client, and vice versa for \code{S->C};
    	\code{f: 50} means that the pair appears in the clustering with a \emph{frequency of 50};
    	\code{Signature notation} shows a summary of 2 sets of 50 instances of packet sequences.
		Example: \code{C->S 556, 1293 f: 50} means that the pair of packets with lengths 556 (client-to-server) 
		and 1293 (server-to-client) appear 50 times in the cluster.
		\label{fig:clustering-signature}}
		\vspace{-0.5em}
\end{figure*}

\subsection{Training}
\label{sect:system-training}
The training component is responsible for the extraction of
packet-level signatures for a device the attacker wants to profile and attack.
It consists of 5 steps (see Figure~\ref{fig:inspecto}). 


\myparagraph{Data Collection.}
\label{sect:data-collection}
The first step towards signature generation is to collect a \emph{training set} 
for the device. 
A training set is a network trace (a PCAP file) that contains the network traffic
generated by the device and smartphone as a result of \events;
\rahmedit{this trace is accompanied by a text file that contains the set of \event timestamps.}

\tool partially automates training set collection by providing a shell script that uses the Android Debug Bridge (\code{adb})~\cite{adbarticle} to issue touch inputs on the smartphone's screen.
\jvedit{The script is run on a laptop that acts as the controller in Figure~\ref{fig:combined-setup}.}
The script is tailored to issue the sequence of touch events corresponding to the \events for which a training set is to be generated.
For example, if a training set is desired for a smart plug's ON and OFF \events, the script issues a touch event at the screen coordinates that correspond to the respective buttons in the user interface of the plug's official Android app.
As device vendors may choose arbitrary positions for the buttons in their respective Android applications, and since the feature sets differ from device to device, the script must be manually modified for the given device.
The script issues the touch sequence corresponding to each specific \event $n$ times, each separated by $m$ 
seconds.\footnote{We selected $m=131$ seconds to allow sufficient time such that there is no 
overlap between events. Section~\ref{sect:parameter-sensitivity} provides more explanation for this choice
with respect to other parameters.}
\rahmedit{
The results reported in this paper use $n=50$ or $n=100$ depending on the \event type 
(see Section~\ref{sect:training-phase}).
}
The script also outputs the 
current timestamp to a file on the laptop when it issues an \event.
\jvedit{To collect a training set, we do the following: (1) start \code{tcpdump} on the router's interfaces; (2) start the script; (3) terminate \code{tcpdump} after the $n$-th event has been issued.} 
This leaves us with a set of PCAP files and \event timestamps, which constitute our raw training set. 
 
We base our signature generation on the traces collected from the 
router's local interfaces as they are the vantage points that provide the most
comprehensive information: they include both local traffic and Internet traffic.
This allows \tool to exhaustively analyze all network packets generated 
in the communications between the device, smartphone, and Internet hosts
on a per device basis.
As signatures are based entirely on packet lengths and 
directions, signatures present in Internet traffic
(\ie \devicecloudcomm and \phonecloudcomm traffic)
are applicable on the WAN side of the router, despite being 
extracted from traces captured within the local network


\myparagraph{Trace Filtering.}
\label{sect:data-preprocessing}
Next, \tool filters the collected raw training set to discard traffic that is unrelated to a user's operation of a smart home device.
All packets, where neither the source nor destination IP matches that of the device or the controlling smartphone, are dropped.
Additionally, all packets that do not lie within a time window \timewindow after each timestamped \event are discarded.
\rahmedit{
We selected $\timewindow=15$ seconds to allow sufficient time for all network traffic related to the event to complete.  
We also performed a sensitivity study that confirmed this was a conservative choice (see Section~\ref{sect:parameter-sensitivity}).
}

\tool next reassembles all TCP connections in the filtered trace.
Given the set of reassembled TCP connections, we now turn our attention 
to the packets $P$ that carry the TCP payload.
For TLS connections, $P$ is limited further 
to only be the subset of packets that are labeled as ``Application Data'' in the unencrypted TLS record header~\cite{tls-rfc}.
By only considering packets in $P$, we ensure that the inherently unpredictable control packets (\eg TCP ACKs and TLS key negotiation) do not become part of the signature as $P$ only contains packets with application layer payload.

We next construct the set $P'$  by forming \emph{packet pairs} from the packets in $P$
(see Definition~\ref{def:packet-pair}).
This is motivated by the following observation:
the deterministic sequence of packets that make up packet-level signatures often stem from a
\emph{request-reply exchange} between the device, smartphones, and some Internet hosts
(see Section~\ref{sect:network-behavior}).
Furthermore, since a packet pair is the simplest possible pattern, and since longer
patterns (\ie packet sequences---see Definition~\ref{def:packet-sequence}) can be 
reconstructed from packet pairs, we look for these packet pairs in the training set.
For the TP-Link plug example in Figure~\ref{fig:inspecto}, \tool reassembles
\code{<..., C-556, S-1293, ...>}, \code{<..., C-237, S-826, ...>},
\etc{} as TCP connections.
Then, \tool extracts \code{<C-556, S-1293>}, \code{<C-237,} \code{S-826>}, \etc{} as packet pairs.

\begin{mdframed}
\theoremstyle{definition}
\begin{definition}{\textbf{\textit{Packet Pair.}}}
\label{def:packet-pair}
\textit{Let \PCspace be the ordered set of packets with 
TCP payload that belong to TCP connection $c$, let \PCispace denote 
the $i$-th packet in \PC, and let $C$ and $S$ each denote
client-to-server and server-to-client packet directions, respectively,
where a client is a smartphone or a device.
A packet pair $p$ is then \Pequals\CPCi\Pcomma\SPCinext\Pclose
or \Pequals\SPCi\Pcomma\CPCinext\Pclose
\emph{iff} \PCispace and \PCinextspace go in opposite directions.
Otherwise, if \PCispace and \PCinextspace go in the same direction,
or if \PCispace is the last packet in \PC, the packet pair 
\Pequals\CPCi\Pcomma\Nil\Pclose or \Pequals\SPCi\Pcomma\Nil\Pclose is formed, 
and packet \PCinext, if any, 
is paired with packet \PCinextnext.}
\end{definition}
\end{mdframed}

\myparagraph{Pair Clustering.}
\label{sect:clustering}
After forming a set of packet pairs, relevant packet pairs (\ie those that consistently occur after an \event) must next be separated from irrelevant ones.
This selection also needs to take into account that the potentially relevant packet pairs may have slight variations in lengths.  
Since we do not know in advance the packet lengths in the pairs, we use an
unsupervised learning algorithm: DBSCAN~\cite{dbscan}.

DBSCAN is provided with a distance function for comparing 
the similarity of two packet pairs, say $p_1$ and $p_2$.
The distance is maximal if the packet directions are different, \eg if $p_1$ 
is comprised of a packet going from a local device to an Internet host
followed by a packet going from an Internet host to a local device, 
while $p_2$ is comprised of a packet going from an Internet host to a local 
device followed by a packet going from a local device to an Internet host.
If the packet directions match, the distance is simply the Euclidean distance 
between the two pairs, \ie $\sqrt{\left( p^1_1 - p^1_2 \right)^2 + \left( p^2_1 - p^2_2 \right)^2}$, 
where $p^i_j$ refers to the packet length of the $i$-th element of pair $j$.
DBSCAN's parameters are $\epsilon$ and $\mathtt{minPts}$,
which specify the neighborhood radius to consider when determining
core points and the minimum number of points in that neighborhood
for a point to become a core point, respectively. We choose
$\epsilon=10$ and $\mathtt{minPts}=\lfloor n-0.1n \rfloor$, where $n$
is the total number of \events. We allow a slack of $0.1n$ to
$\mathtt{minPts}$ to take into account that event-related traffic could
occasionally have missing pairs, for example caused by the phone app
not responding to some of the automated \events.
We study the sensitivity of \tool parameter values in Section~\ref{sect:parameter-sensitivity}.

Figure~\ref{fig:clustering-signature}(a) illustrates the pair 
clustering process for  TP-Link plug. 
There are 50 ON and 50 OFF actions, and there must be at least 45 ($n=50 \implies \mathtt{minPts}=\lfloor 50-0.1 \times 50 \rfloor=45$) similar packet pairs to form a cluster. 
Two clusters are formed among the data points,  \ie those with frequencies \code{f: 50} and \code{f: 98}, respectively.
Since these two clusters contain similar packet pairs that occur during \timewindow, this indicates with high confidence that the packets are related to the \event.

\myparagraph{Signature Creation.}
\label{sect:post-processing}
Given the output produced by DBSCAN, \tool next drops all clusters whose frequencies are not in the interval $\left[ \lfloor n-0.1n \rfloor , \lceil n + 0.1n \rceil \right]$ in order to only include in the signature those clusters whose frequencies align closely with the number of \events $n$.
Intuitively, this step is to deal with \emph{chatty devices}, namely 
devices that communicate continuously/periodically while not generating 
\events.
Consequently, \tool only picks the cluster \code{Pairs 1} with frequency 50 for the TP-Link plug example in Figure~\ref{fig:clustering-signature} as a signature candidate since 50 is in $\left[ \lfloor n-0.1n \rfloor , \lceil n + 0.1n \rceil \right]=\left[ 45, 55 \right]$ when $n=50$, whereas 98 is not.
As a pair from this cluster occurs exactly once during \timewindow, there is high confidence that the pair is related to the \event.

\tool{} next attempts to concatenate packet pairs in the clusters so as to reassemble the longest 
\emph{packet sequences} possible (see Definition~\ref{def:packet-sequence}),
which increases the odds that a signature is unique.
Naturally, packet pair concatenation is only performed when a device has more than one cluster. 
This is the case for the \arlocamera, but not the \tplinkplug.
Packet pairs in clusters $x$ and $y$ are concatenated \textit{iff} 
for each packet pair $p_x$ in $x$, there exists a packet pair $p_y$ in $y$ such 
that $p_x$ and $p_y$ occurred consecutively in the same TCP connection.
If there are more pairs in $y$ than in $x$, the extra pairs of $y$ are simply 
dropped. 
The result is referred to as a \emph{set of packet sequences}  
(see Definition~\ref{def:set-packet-sequences}) and is considered for further concatenation with other clusters if possible.

\begin{mdframed}
\theoremstyle{definition}
\begin{definition}{\textbf{\textit{Packet Sequence.}}}
\label{def:packet-sequence}
\textit{A packet sequence $s$ is formed by
joining  packet pairs $p_1$ and $p_2$ \emph{iff} $p_1$ and $p_2$ are
both in $P_c$ (same TCP connection) \emph{and} the packets in $p_1$ occur immediately before
the packets in $p_2$ in $P_c$. Note that the packet sequence $s$ resulting
from joining $p_1$ and $p_2$ can be of length 2, 3, or 4, depending
on whether or not the second element of $p_1$ and/or $p_2$ 
is nil.}
\end{definition}
\end{mdframed}

\begin{mdframed}
\theoremstyle{definition}
\begin{definition}{\textbf{\textit{Set of Packet Sequences.}}}
\label{def:set-packet-sequences}
\textit{A set of packet sequences $S$ is a
set of similar packet sequences. Two packet sequences $s_1$ and $s_2$
are similar and thus belong to the same set $S$ 
\emph{iff} they (1) contain the same number of 
packets, (2) the packets at corresponding indices of $s_1$ and $s_2$
go in the same direction, and (3) the Euclidean distance between the packet 
lengths at corresponding indices of $s_1$ and $s_2$ is below a 
threshold---packet lengths in packet sequences inherit the slight variations that stem from 
packet pairs.}
\end{definition}
\end{mdframed}

Figure~\ref{fig:clustering-signature}(b) shows how pair clustering
produces 3 clusters around the pairs \code{<C-339, S-329>} (\ie cluster \code{Pairs 1}),
\code{<C-[364-365], S-[1061-1070]>} (\ie cluster \code{Pairs 2}), and 
\code{<C-[271-273], S-[499-505]>} (\ie cluster \code{Pairs 3}) for 
the \arlocamera. The notation \code{C-[$l_1-l_2$]} or \code{S-[$l_1-l_2$]} 
indicates that the packet length may vary in the range between $l_1$ and $l_2$.
Each pair from cluster \code{Pairs 1} and 
each pair from cluster \code{Pairs 2} are then 
concatenated into a sequence in \code{Sequences 1} (a set of packet sequences)
as they appear consecutively in the same TCP connection, 
\ie \code{Pair 1.1} with \code{Pair 2.1}, \code{Pair 1.2} with \code{Pair 2.2}, ...,
\code{Pair 1.50} with \code{Pair 2.50}.
The cluster \code{Pairs 3} is finalized as the set \code{Sequences 2} as its members appear in different TCP connections than the members of \code{Sequences 1}.
Thus, the initial 3 clusters of packet pairs are reduced to 2 sets of packet sequences.
For the TP-Link plug, no concatenation is performed since there is only a single cluster, \code{Pairs 1}, which is finalized as the set \code{Sequences 1}.


Finally, \tool{} sorts the sets of packet sequences based on the timing of the sets' members to form a \emph{list of packet sequence sets} (see Definition~\ref{def:list-packet-sequence-sets}).
For example, for the \arlocamera, this step produces a list 
in which the set \code{Sequences 1} precedes the set \code{Sequences 2}
because there is always a packet sequence in \code{Sequences 1} that precedes a packet sequence in 
\code{Sequences 2}.
The purpose of this step is to make the temporal order of the sets of packet sequences part of the final signature.
If no such order can be established, the set with the shorter packet sequences is discarded.
Manual inspection of some devices suggests that the earlier sequence 
will often be the control command sent from an Internet host followed 
by the device's acknowledgment of the command, while the 
later sequence will stem from the device initiating communication 
with some other Internet host to inform that host
about its change in status.

\begin{mdframed}
\theoremstyle{definition}
\begin{definition}{\textbf{\textit{List of Packet Sequence Sets.}}}
\label{def:list-packet-sequence-sets}
\textit{A list of packet sequence sets is a list that contains sets of packet 
sequences that are sorted based on the occurrence of the set members
in time.
Set $S_x$ goes before set $S_y$ \emph{iff} for 
each sequence $s_x$ in $S_x$, there exists a sequence $s_y$ in $S_y$ 
that occurred after $s_x$ within \timewindow.}
\end{definition}
\end{mdframed}

\myparagraph{Signature Validation.}
Before finalizing the signature, we validate it by running the
detection algorithm (see Section~\ref{sect:system-detection}) against the raw training set
that was used to generate the signature.
If \tool detects at most
$n$ \events, and the timestamps of detected \events match the timestamps for \events recorded during training, the signature is finalized as a valid \emph{packet-level signature} (see Definition~\ref{def:packet-level-signature}) 
and stored in a signature file.
A signature can fail this check if it detects more events than the actual
number of \events in the training set (\ie false positives). 
This can happen if the packet sequences in the signature frequently appear outside \timewindow.

\begin{mdframed}
\theoremstyle{definition}
\begin{definition}{\textbf{\textit{Packet-level Signature.}}}
\label{def:packet-level-signature}
\textit{A packet-level signature is then a list of packet sequence sets that has been
validated and finalized.}
\end{definition}
\end{mdframed}

\myparagraph{Signature File.} 
A signature file stores a packet-level signature.
Figure~\ref{fig:clustering-signature} shows that the \tplinkplug signature consists of 50 instances of packet sequences in set \code{Sequences 1}, but only one instance will be used during detection since all 50 are identical.
Figure~\ref{fig:clustering-signature}(b) shows the signature file (on the right-hand side) for the \arlocamera. 
It is a list that orders the two sets of packet sequences, \code{Sequences 1} and \code{Sequences 2}.
\code{Sequences 1} is comprised of 50 packet sequences, each comprised of two packet pairs.
\code{Sequences 2} is comprised of another 50 packet sequences, each comprised of a single packet pair.
Since the sequences vary slightly in each set, all unique variations are considered during detection.


\subsection{Detection\label{sect:system-detection}}
For signature detection, \tool treats a network
trace as a stream of packets and presents each individual packet to a set of state machines. 
A state machine is maintained for each packet sequence of the signature for
each flow, i.e., TCP connection for the WAN sniffer or
layer-2 flow for the Wi-Fi sniffer. A packet
is only presented to the state machines associated with
the flow that the packet pertains to. A state machine advances to its next state if the 
packet matches the next packet (in terms of length and direction) in the modeled packet sequence. The 
state machines respond differently to packets that do not match the expected 
next packet depending on whether detection is applied at layer-2 or layer-3.
For layer-2, such packets are simply ignored, whereas for
layer-3 such packets cause the state machine to discard
the current partial match. When a state
machine reaches its terminal state, the packet sequence
match is reported to a secondary module. This module
waits for a packet sequence match for each packet sequence of the signature 
and verifies the inter-sequence
timing constraints before finally declaring a signature
match.
Please see \appdx{Appendix~\ref{sect:system-detection-cr}} for a more detailed explanation of the detection.

\begin{table*}[htb!]
  \centering
  \begin{center}
  { \footnotesize
  \begin{tabular}{| p{22mm} | p{18mm} | c | c | c | c | c | c | c |}
    \hline
        \textbf{Device} & \textbf{Event} & \textbf{Signature} & \textbf{Communication} & \textbf{Duration (ms)} & \multicolumn{4}{ c |}{\textbf{Matching (Per 100 Events)}}\\
	\cline{6-9}
        & & & & \textbf{Min./Avg./Max.} & \textbf{WAN} & \textbf{FPR} & \textbf{Wi-Fi} & \textbf{FPR}\\
        & & & & & \textbf{Snif.} & & \textbf{Snif.} & \\
    \hline
    	\multicolumn{9}{| c |}{\textbf{Plugs}}\\
	\hline
        Amazon plug & ON & \textbf{S1:} S-[443-445] & \devicecloudcomm & 1,232 / 2,465 / 4,537 & 98 & 0 & 99 & 0 \\
        & & \textbf{S2:} C-1099 S-235 & & & & & & \\
	\cline{2-3}
        & OFF & \textbf{S1:} S-[444-446] & & & & & & \\
        & & \textbf{S2:} C-1179 S-235 & & & & & & \\
        & & \textbf{S3:} C-1514 C-103 S-235 & & & & & & \\
	\hline
        WeMo plug & ON/OFF & \textbf{S1:} PH-259 PH-475 D-246 & \phonedevicecomm & 33 / 42 / 134 & - & - & 100 & 0 \\
        & & & & & & & & \\
	\hline
        WeMo Insight plug & ON/OFF & \textbf{S1:} PH-259 PH-475 D-246 & \phonedevicecomm & 32 / 39 / 97 & - & - & 99 & 0 \\
        & & & & & & & & \\
	\hline
        TP-Link plug & ON & \textbf{S1:} C-556 S-1293 & \devicecloudcomm & 75 / 85 / 204 & 99 & 0 & - & - \\
    \cline{2-3}
        & OFF & \textbf{S1:} C-557 S-[1294-1295] & & & & & & \\
    \cline{2-9}
        & ON & \textbf{S1:} PH-112 D-115 & \phonedevicecomm & 225 / 325 / 3,328 & - & - & 99 & 0 \\
        & & \textbf{S2:} C-556 S-1293 & \& & & & & & \\
	\cline{2-3}
	     & ON & \textbf{S1:} PH-112 D-115 & \devicecloudcomm & & & & & \\
        & & \textbf{S2:} C-557 S-[1294-1295] & & & & & & \\
	\hline
        D-Link plug & ON/OFF & \textbf{S1:} S-91 S-1227 C-784 & \devicecloudcomm & 4 / 1,194 / 8,060 & 95 & 0 & 95 & 0 \\
        & & \textbf{S2:} C-1052 S-647 & & & & & & \\
    \cline{2-9}
    	& ON & \textbf{S1:} C-[1109-1123] S-613 & \phonecloudcomm & 35 / 41 / 176 & 98 & 0 & 98 & 0 \\
    \cline{2-3}
        & OFF & \textbf{S1:} C-[1110-1124] S-613 & & & & & & \\
    \hline
        SmartThings plug & ON & \textbf{S1:} C-699 S-511 & \phonecloudcomm & 335 / 537 / 2,223 & 92 & 0 & 92 & 0 \\
        & & \textbf{S2:} S-777 C-136 & & & & & & \\
	\cline{2-3}
        & OFF & \textbf{S1:} C-700 S-511 & & & & & & \\
        & & \textbf{S2:} S-780 C-136 & & & & & & \\
    \hline
    	\multicolumn{9}{| c |}{\textbf{Light Bulbs}}\\
	\hline
        Sengled light bulb & ON & \textbf{S1:} S-[217-218] C-[209-210] & \devicecloudcomm & 4,304 / 6,238 / 8,145 & 97 & 0 & - & - \\
        & & \textbf{S2:} C-430 & & & & & & \\
        & & \textbf{S3:} C-466 & & & & & & \\
	\cline{2-3}
        & OFF & \textbf{S1:} S-[217-218] C-[209-210] & & & & & & \\
        & & \textbf{S2:} C-430 & & & & & & \\
        & & \textbf{S3:} C-465 & & & & & & \\
	\cline{2-9}
        & ON & \textbf{S1:} C-211 S-1063 & \phonecloudcomm & 4,375 / 6,356 / 9,132 & 93 & 0 & 97 & 0 \\
        & & \textbf{S2:} S-1277 & & & & & & \\
	\cline{2-3}
        & OFF & \textbf{S1:} C-211 S-1063 S-1276 & & & & & & \\
	\cline{2-9}
        & Intensity & \textbf{S1:} S-[216-220] & \devicecloudcomm & 16 / 74 / 824 & 99 & 2 & - & - \\
        & & C-[208-210] & & & & & & \\
	\cline{2-9}
        & Intensity & \textbf{S1:} C-[215-217] & \phonecloudcomm & 3,916 / 5,573 / 7,171 & 99 & 0 & 99 & 0 \\
        & & S-[1275-1277] & & & & & & \\
	\hline
        Hue light bulb & ON & \textbf{S1:} C-364 & \devicecloudcomm & 11,019 / 12,787 / & - & - & - & - \\
        & & \textbf{S2:} D-88 & \& & 14,353 & & & & \\
	\cline{2-3}
        & OFF & \textbf{S1:} C-365 & \phonedevicecomm & & & & & \\
		& & \textbf{S2:} D-88 & & & & & & \\
	\hline
        TP-Link light bulb & ON & \textbf{S1:} PH-198 D-227 & \phonedevicecomm & 8 / 77 / 148 & - & - & 100 & 4 \\
	\cline{2-3}
        & OFF & \textbf{S1:} PH-198 D-244 & & & & & & \\
    \cline{2-9}
        & Intensity & \textbf{S1:} PH-[240-242] D-[287-289] & \phonedevicecomm & 7 / 84 / 212 & - & - & 100 & 0 \\
    \cline{2-9}
        & Color & \textbf{S1:} PH-317 D-287 & \phonedevicecomm & 6 / 89 / 174 & - & - & 100 & 0 \\
    \hline
    	\multicolumn{9}{| c |}{\textbf{Thermostats}}\\
	\hline
    	Nest thermostat & Fan ON & \textbf{S1:} C-[891-894] S-[830-834] & \phonecloudcomm & 91 / 111 / 1,072 & 93 & 0 & 93 & 1 \\
    \cline{2-3}
    	& Fan OFF & \textbf{S1:} C-[858-860] S-[829-834] & & & & & & \\
   	\hline
        Ecobee thermostat & HVAC Auto & \textbf{S1:} S-1300 C-640 & \phonecloudcomm & 121 / 229 / 667 & 100 & 0 & 99 & 0 \\
	\cline{2-3}
        & HVAC OFF & \textbf{S1:} C-1299 C-640 & & & & & & \\
	\cline{2-9}
        & Fan ON & \textbf{S1:} S-1387 C-640 & \phonecloudcomm & 117 / 232 / 1,776 & 100 & 0 & 100 & 0 \\
	\cline{2-3}
        & Fan Auto & \textbf{S1:} C-1389 C-640 & & & & & & \\
    \hline
    	\multicolumn{9}{| c |}{\textbf{Sprinklers}}\\
	\hline
        Rachio sprinkler & Quick Run & \textbf{S1:} S-267 C-155 & \devicecloudcomm & 1,972 / 2,180 / 2,450 & 100 & 0 & 100 & 0 \\
	\cline{2-3}
        & Stop & \textbf{S1:} C-496 C-155 C-395 & & & & & & \\
	\cline{2-9}
        & Standby/Active & \textbf{S1:} S-299 C-155 C-395 & \devicecloudcomm & 276 / 690 / 2,538 & 100 & 0 & 100 & 0 \\
    \hline
        Blossom sprinkler & Quick Run & \textbf{S1:} C-326 & \devicecloudcomm & 701 / 3,470 / 8,431 & 96 & 0 & 96 & 0 \\
        & & \textbf{S2:} C-177 S-505 & & & & & & \\
	\cline{2-3}
        & Stop & \textbf{S1:} C-326 & & & & & & \\
        & & \textbf{S2:} C-177 S-458 & & & & & & \\
        & & \textbf{S3:} C-238 C-56 S-388 & & & & & & \\
    \cline{2-9}
        & Quick Run & \textbf{S1:} C-649 S-459 C-574 S-507 & \phonecloudcomm & 70 / 956 / 3,337 & 93 & 0 & 93 & 0 \\
        & & \textbf{S2:} S-[135-139] & & & & & & \\
	\cline{2-3}
        & Stop & \textbf{S1:} C-617 S-431 & & & & & & \\
    \cline{2-9}
        & Hibernate & \textbf{S1:} C-621 S-493 & \phonecloudcomm & 121 / 494 / 1,798 & 95 & 0 & 93 & 0 \\
	\cline{2-3}
        & Active & \textbf{S1:} C-622 S-494 & & & & & & \\
        & & \textbf{S2:} S-599 C-566 S-554 C-566 & & & & & & \\
    \hline
  \end{tabular}
  }
  \end{center}
\end{table*}

\begin{table*}[htb!]
  \centering
  \begin{center}
  { \footnotesize
  \begin{tabular}{| p{25mm} | p{20mm} | c | c | c | c | c | c | c |}
    \hline
        \textbf{Device} & \textbf{Event} & \textbf{Signature} & \textbf{Communication} & \textbf{Duration (ms)} & \multicolumn{4}{ c |}{\textbf{Matching (Per 100 Events)}}\\
	\cline{6-9}
        & & & & \textbf{Min./Avg./Max.} & \textbf{WAN} & \textbf{FPR} & \textbf{Wi-Fi} & \textbf{FPR}\\
        & & & & & \textbf{Snif.} & & \textbf{Snif.} & \\
    \hline
    \multicolumn{9}{| c |}{\textbf{Home Security Devices}}\\

		\hline
        Ring alarm & Arm & \textbf{S1:} S-99 S-254 C-99 & \devicecloudcomm & 275 / 410 / 605 & 98 & 0 & 95 & 0 \\
        & & S-[181-183] C-99 & & & & & & \\
	\cline{2-3}
        & Disarm & \textbf{S1:} S-99 S-255 C-99 & & & & & & \\
        & & S-[181-183] C-99 & & & & & & \\
    \hline
        Arlo camera & Stream ON & \textbf{S1:} C-[338-339] S-[326-329] & \phonecloudcomm & 46 / 78 / 194 & 99 & 2 & 98 & 3 \\
        & & C-[364-365] S-[1061-1070] & & & & & & \\
        & & \textbf{S2:} C-[271-273] S-[499-505] & & & & & & \\
	\cline{2-3}
        & Stream OFF & \textbf{S1:} C-[445-449] S-442 & & & & & & \\
	\hline
        D-Link siren & ON & \textbf{S1:} C-1076 S-593 & \phonecloudcomm & 36 / 37 / 65 & 100 & 0 & 98 & 0 \\
	\cline{2-3}
        & OFF & \textbf{S1:} C-1023 S-613 & & & & & & \\
	\hline
        Kwikset door lock & Lock & \textbf{S1:} C-699 S-511 & \phonecloudcomm & 173 / 395 / 2,874 & 100 & 0 & 100 & 0 \\
        & & \textbf{S2:} S-639 C-136 & & & & & & \\
	\cline{2-3}
        & Unlock & \textbf{S1:} C-701 S-511 & & & & & & \\
        & & \textbf{S2:} S-647 C-136 & & & & & & \\
    \hline
    	\multicolumn{9}{| c |}{\textbf{Others}}\\
	\hline
        Roomba robot & Clean & \textbf{S1:} S-[1014-1015] C-105 & \phonecloudcomm & 123 / 2,038 / 5,418 & 91 & 0 & 94 & 0 \\
        & & S-432 C-105 & & & & & & \\
	\cline{2-3}
        & Back-to-station & \textbf{S1:} S-440 C-105 & & & & & & \\
        & & S-[1018-1024] C-105 & & & & & & \\
    \hline
    	\multicolumn{5}{| r |}{Average} & 97.05 & 0.18 & 97.48 & 0.32 \\
    \hline
  \end{tabular}
  }
  \end{center}
	\caption{Smart home devices found to exhibit \phonecloudcomm, \devicecloudcomm, and \phonedevicecomm signatures.
  	Prefix PH indicates Phone-to-device direction and prefix D indicates Device-to-phone direction in Signature column.
  	\label{tab:summary-signatures}}
	
		\vspace{-2em}
\end{table*}

\section{Evaluation}
\label{sect:evaluation}
In this section, we present the evaluation of  \tool.
In  Section~\ref{sect:training-phase}, we show that  \tool automatically extracted event signatures for 18 devices as summarized in Table~\ref{tab:summary-signatures}---11 of which are
the most popular devices on Amazon (see Table~\ref{tab:smarthome-devices}). 
In Section~\ref{sect:detection-phase}, we used the extracted signatures  to detect \events in a trace collected from a realistic experiment on our smart home testbed.
Section~\ref{sect:negative-control} discusses the results of negative control experiments: it demonstrates the uniqueness of the \tool signatures in large  (\ie with hundreds of millions of packets), publicly available, packet traces from smart home and  office environments.
\rahmedit{
\majorrevision{
Section~\ref{sect:home-automation} discusses the results of our experiments when
devices are triggered remotely from a smartphone and via a home automation service.
Section~\ref{sect:same-vendor} shows the uniqueness of signatures for devices from the same vendor.
Section~\ref{sect:public-dataset} discusses our findings when we used \tool{} to extract signatures from a public dataset~\cite{ren2019information}.
Finally, Section~\ref{sect:parameter-sensitivity} discusses the selection and sensitivity of the parameters used to extract signatures.}
}

\subsection{Extracting Signatures from Smart Home Devices}
\label{sect:training-phase}
\myparagraph{Training Dataset.} In order to evaluate the generalizability of packet-level signatures, we first
used \tool to automate the collection of training sets (see 
Section~\ref{sect:data-collection}) for all 19 smart home devices 
(see Table~\ref{tab:smarthome-devices}). 
Training sets were collected for every device under test, individually 
without any background traffic (see Figure~\ref{fig:combined-setup}).
The automation script generated a total of 100 \events 
for the device.
For \events with binary values, the script generated
$n=50$ events for each event type (\eg 50 ON and 50 OFF events).
For \events with continuous values, the script generated
$n=100$ events (\eg 100 intensity events for the \sengledbulb).

\myparagraph{Results Summary.} For each training set, we used \tool to extract packet-level signatures 
(see Section~\ref{sect:system-training}) for each event type 
of the respective device. 
In summary, \tool extracted signatures from 18 devices
(see Table~\ref{tab:summary-signatures}).
\majorrevision{
The signatures span a wide range of \event types:
binary (\eg ON/OFF) and non-binary (\eg light bulb intensity, color, etc.).
Similar to our manual observation described in Section~\ref{sect:network-behavior}, 
we again see that these events are identifiable by the request-reply pattern.
}

Table~\ref{tab:summary-signatures} 
presents the signatures that \tool identified.\techreport{\footnote{
\phonedevicecomm signatures are observable only by the Wi-Fi sniffer.
The \sengledbulb's \devicecloudcomm signatures are sent by the Zigbee hub to the cloud 
through the Ethernet interface; thus, they are not 	observable by the \wifisniffer.
The \huebulb has unique signatures; they consist of a pair, in which
one is a \devicecloudcomm packet coming from the Zigbee hub to the cloud---this 
is observable only by the \wansniffer since the hub is an Ethernet device---and 
the other one is a \phonecloudcomm packet---this is observable only by the \wifisniffer;
thus we did not use the signatures to perform detection since they partially belong to both adversaries.
\label{footnote:main-table-notes}
}}
Each line in a signature cell represents 
a packet sequence set, and the vertical positioning of these lines reflects
the ordering of the packet sequence sets in the signature (see 
Section~\ref{sect:system-training} for the notation).

\rahmedit{
\tool performed well in extracting signatures: it has successfully extracted packet-level signatures
that are observable in the device's \phonecloudcomm, \devicecloudcomm, and 
\phonedevicecomm communications (see Table~\ref{tab:summary-signatures}).
Although the traffic is typically encrypted
using TLSv1.2,  the \event still manifests itself in the form of a packet-level 
signature in the \phonecloudcomm or \devicecloudcomm communication.
\tool also extracted signatures from the \phonedevicecomm communication
for some of the devices.
These signatures are extracted typically from unencrypted local 
TCP/HTTP communication between the smartphone and the device. 
}

\mysubparagraph{Smart Plugs.}
\tool extracted signatures from all 6 plugs: the Amazon, WeMo, WeMo Insight, TP-Link, D-Link, and SmartThings plugs.
The Amazon, D-Link, and SmartThings plugs
have signatures in
the \phonecloudcomm or \devicecloudcomm communication, or both.
The \tplinkplug has signatures in both the \devicecloudcomm and \phonedevicecomm
communications.
Both the WeMo and WeMo Insight plugs have signatures in the \phonedevicecomm
communication.
In general, the signatures allow us to differentiate ON from OFF
except for the WeMo, WeMo Insight, TP-Link plugs'
\phonedevicecomm
communication, and \dlinkplug's \devicecloudcomm communication
(see Table~\ref{tab:summary-signatures}).

\mysubparagraph{Light Bulbs.}
\tool extracted signatures from 3 light bulbs: the Sengled, Hue, and TP-Link light bulbs.
The \sengledbulb has signatures in both the \phonecloudcomm and \devicecloudcomm communications.
\rahmedit{
The \huebulb has signatures in both \devicecloudcomm and \phonedevicecomm
communications.
}
The TP-Link light bulb has signatures only in the \phonedevicecomm
communication.
Table~\ref{tab:summary-signatures} shows that \tool also extracted signatures
for events other than ON and OFF: Intensity and Color. 

\mysubparagraph{Thermostats.}
\tool extracted signatures for both the Nest and Ecobee thermostats.
Both thermostats have \phonecloudcomm signatures.
The signatures allow us to differentiate
Fan ON/OFF/Auto events. The \ecobeethermostat's signatures also leak
information about its HVAC Auto/OFF events.

\mysubparagraph{Sprinklers.}
\tool extracted signatures from both the \rachiosprinkler and
\blossomsprinkler.
Both sprinklers have signatures in both the \devicecloudcomm and \phonecloudcomm
communications.
The signatures allow us to differentiate
Quick Run/Stop and Standby/Hibernate/Active events.

\mysubparagraph{Home Security Devices.}
A highlight  is that \tool 
extracted signatures from home security devices.
Notably, the \ringalarm has signatures that allow us
to differentiate Arm/Disarm events in the \devicecloudcomm communication.
The \arlocamera has signatures for Stream ON/OFF events,
the \dlinksiren for ON/OFF events, and
the \kwiksetdoorlock for Lock/Unlock events in the \phonecloudcomm communication.

\mysubparagraph{\roombarobot.}
Finally, \tool also extracted signatures from the \roombarobot in the \phonecloudcomm communication.
These signatures allow us to differentiate Clean/Back-to-station events.

\myparagraph{Signature Validity.\label{sect:signature-validity}}
Recall that signature validation rejects a signature candidate whose sequences 
are present not only in the time window \timewindow, 
but also during the subsequent idle period 
(see Section~\ref{sect:system-training}).
\brianedit{We saw such a signature candidate for one device, namely the \lifxbulb.
\tool captured a signature candidate that is present also in the idle period
of the TCP communication and then rejected the signature during the validation phase.
Manual inspection revealed that the \lifxbulb uses unidirectional 
UDP communication (\ie no request-reply pattern) for \events.}

\mycomment{
\myparagraph{Signature Validity.\label{sect:signature-validity}}
When \tool has a candidate signature, it performs
\emph{signature validation} before it finalizes the candidate as a signature;
see Section~\ref{sect:system-training} for details.
Recall that signature validation filters the
signature candidates that are not sufficiently unique to represent an event.
This can for example be the case when the signature is generated from 
traffic that happens to be present in the time window \timewindow, 
but which also occurs during the subsequent idle period 
(see Section~\ref{sect:system-training}). 
We saw its manifestation for one device, namely the \lifxbulb.
Manual inspection revealed that it uses unidirectional 
UDP communication for \events---it does not consist of request-reply packet 
pairs since it is a best-effort service.
Since \tool focuses on TCP communication, the previously captured signature
candidate turned out to be communication for other purposes. 
We then drop such a signature candidate.

In some cases, the validation does not capture all 100 events.
Manual inspection revealed that in these few cases, the signatures were 
only partially present or they are present with more variations, \eg
\code{S2: S-777 C-136} for the SmartThings plug might become
\code{S2: S-612 C-136} in some occasions.
}

\subsection{Smart Home Testbed Experiment}
\label{sect:detection-phase}
\myparagraph{Testing Dataset.}
To evaluate the effectiveness of packet-level signatures  in detecting \events,
we collected a separate set of network traces and used \tool to perform detection 
on them.
We used the setup in 
Section~\ref{sect:experimental-testbed} to collect one dataset for every device.
Our smart home setup consists of 13 of the smart home devices
presented in Table~\ref{tab:smarthome-devices}: the \wemoplug, \wemoinsightplug,
\huebulb, \lifxbulb, \nestthermostat, \arlocamera, \tplinkplug, \dlinkplug,
\dlinksiren, \tplinkbulb, \smartthingsplug, \blossomsprinkler, and
\kwiksetdoorlock. This fixed set of 13 devices was our initial 
setup---it gives us the flexibility to 
test additional devices without changing the smart home setup and needing to rerun all the experiments,
yet still includes a variety of devices that generate background traffic.
While collecting a dataset, we triggered \events for the device under test. 
\rahmedit{
At the same time, we also connected the other 12 devices and turned them ON
before we started the experiment---this allows the other devices to generate
network traffic as they communicate with their cloud servers. 
However, we did not trigger \events for these other devices.
For the other 6 devices (the \amazonplug,
\sengledbulb, \ecobeethermostat, \rachiosprinkler, \roombarobot, and 
\ringalarm), we triggered \events for the device under test
while having all the 13 devices turned on. 
}
To generate additional background traffic as depicted in Figure~\ref{fig:combined-setup},
we set up 3 general purpose devices: a Motorola Moto g$^6$ phone that would play
a YouTube video playlist, a Google Nexus 5 phone that would play a
Spotify song playlist, and an Apple MacBook Air that would randomly
browse top 10 websites~\cite{topwebsites} every 10-500 seconds.
We used this setup to emulate the network traffic from a smart
home with many active devices.

\myparagraph{Results Summary.}
Table~\ref{tab:summary-signatures} presents the summary of our results (see column ``\textbf{Matching}'').
We collected a dataset with 100 events for every type of \event---for binary \events (\eg ON/OFF), we triggered 50 for each value.
We performed the detection for both 
the \wansniffer{} and \wifisniffer{} adversaries.
For both adversaries, we have a negligible False Positive Rate (FPR) of 0.25 (0.18 for the \wansniffer and 0.32 for the \wifisniffer) per 100 \events for every \event type.

\subsection{Negative Control Experiment}
\label{sect:negative-control}
If the packet-level signatures are to be used to detect \events{} in traffic in the wild, they must be sufficiently unique compared to other traffic to avoid generating false positives.
We evaluated the uniqueness of the signatures by performing signature 
detection on 3 datasets. 
The first 2 datasets serve to evaluate the uniqueness of the signatures among traffic generated by similar devices (\ie other smart 
home devices), while the third dataset serves to evaluate the uniqueness 
of the signatures among traffic generated by general purpose computing devices.

\myparagraph{Dataset 1: UNSW Smart Home Traffic Dataset.}
The first dataset~\cite{sivanathan-classifier} contains network traces for 
26 smart home devices that are \emph{different} from the devices that 
we generated signatures for. 
The list can be found in~\cite{devicelist}. 
The dataset is a collection of 19 PCAP files, with a total size of 12.5GB and a total of 23,013,502 packets.

\myparagraph{Dataset 2: YourThings Smart Home Traffic Dataset.}
The second dataset~\cite{sok-alrawi,yourthings} contains network traces for 45
smart home devices. 
The dataset is a collection of 2,880 PCAP files, with a total size of 270.3GB and 407,851,830 
packets.
There are 3 common devices present in both YourThings and 
our set of 18 devices: the \wemoplug, \roombarobot, and \tplinkbulb. 

\myparagraph{\majorrevision{Dataset 3:} UNB Simulated Office-Space Traffic Dataset.}
The third dataset is the Monday trace of the CICIDS2017 
dataset~\cite{sharafaldin2018toward}. It contains simulated network traffic 
for an office space with two servers and 10 laptops/desktops with diverse 
operating systems.
The dataset we used is a single PCAP file of 10.82GB, with a total of 11,709,971 
packets observed at the WAN interface.

\myparagraph{False Positives.}
For datasets 1 and 3, we performed signature detection for all devices. 
For dataset 2, we only performed signature detection for the 15 of our devices that are {\em not} present in YourThings to avoid the potential for true positives.
We used \wansniffer detection for devices with \phonecloudcomm and \devicecloudcomm signatures, and \wifisniffer detection for all devices.

\mysubparagraph{WAN Sniffer.}
There were no false positives 
across 23,013,502 packets in dataset~1, 
1 false positive for the \sengledbulb across 407,851,830 packets in dataset~2, and
1 false positive for the \nestthermostat across 11,709,971 packets in dataset~3.

\mysubparagraph{Wi-Fi Sniffer.}
\tool detected some false positives due to its
more relaxed matching strategy 
(see Section~\ref{sect:system-detection}).
The results show that the extracted packet-level signatures are unique:
the average FPR is 11 false positives per signature across a total of 442,575,303 packets
from all three datasets (\ie an average of 1 false positive per 40 million packets). 

Further analysis showed that signatures comprised of a single packet pair 
(\eg the \dlinkplug's \phonecloudcomm signatures that only have one request and one reply packet)
contributed the most to the average
FPR---FPR is primarily impacted by signature length, not device type.
Five 3-packet signatures generated 5, 7, 16, 26, and 33 false positives, while
one 4-packet signature generated 2 false positives.
There were also three outliers: two 4-packet signatures generated 46 and 33 false positives,
and a 6-packet signature generated 18 false positives.
This anomaly was due to \tool using the range-based matching strategy 
for these signatures (\appdx{see Appendix~\ref{sect:system-detection-cr}}).
Furthermore, the average of the packet lengths for the signatures that generated false positives
is less than 600 bytes: the packet lengths distribution for our negative datasets shows that
there are significantly more shorter packets than longer packets.

\subsection{Events Triggered Remotely}
\label{sect:home-automation}

Our main dataset, collected using our testbed (see Section~\ref{sect:training-phase}), contains \events triggered by a smartphone that is part of the local network (\ie the smart home testbed). 
However, smart home devices can also be controlled remotely, using home automation frameworks or a remote smartphone. In this section, we summarize our results for these scenarios. Please \appdx{see Appendix~\ref{sect:home-automation-cr}} for details.

\myparagraph{Home Automation Experiment (\ifttt).}
We integrated \ifttt into our existing infrastructure for triggering device events.
\ifttt provides support for 13 out of our 18 devices: no support was  provided at the time of the experiment 
for the \amazonplug, \blossomsprinkler, \roombarobot, \ringalarm, and \nestthermostat.
The main finding is that, from the supported 13 devices, \tool{} successfully extracted \devicecloudcomm signatures 
for 9 devices and 12 \event types.

\myparagraph{Comparison of \devicecloudcomm Signatures.}
We also compared the \devicecloudcomm signatures of 
the \tplinkplug, the \dlinkplug, and the \rachiosprinkler.
Our results show that the majority of \devicecloudcomm signatures are the same or very similar 
across 3 different ways of triggering the devices: local smartphone, remote smartphone, and \ifttt.

\subsection{Devices from the Same Vendor}
\label{sect:same-vendor}

Since the signatures reflect protocol behavior, a natural question to ask is whether devices from the same vendor, which probably run similar protocols, have the same signature. 
In our testbed experiment, we had already extracted signatures from 2 TP-Link devices: the \tplinkplug
and \tplinkbulb (see Table~\ref{tab:summary-signatures}). 
{
	We also acquired, and experimented with, 4 additional devices from TP-Link.
	We defer the detailed results to \appdx{Table~\ref{tab:same-vendor}}.
	In summary, we found that packet-level signatures have some similarities (\eg the \tplinktwooutletplug and \tplinkpowerstrip have similar functionality and have packet lengths 1412B and 88B).
	However, they  are still distinct across different device models and event types, even for devices with similar functionality 
	(\eg the \tplinkplug, \tplinktwooutletplug, and \tplinkpowerstrip).
}


\mycomment{
We present the results from TP-Link devices:
(1) different signatures are extracted from these devices from the same vendor;
(2) devices that are more similar exhibit signatures that are more similar too, but they are still different.
}

\majorrevision{
\subsection{Public Dataset Experiment\label{sec:public-dataset}}

\begin{table*}[t!]
  \centering
  \majorrevision{
  \begin{center}
  { \footnotesize
  \begin{tabular}{| p{25mm} | p{20mm} | c | c |}
    \hline
        \textbf{Device} & \textbf{Event} & \textbf{Signature} & \textbf{Duration (ms)}\\
	\hline
		\multicolumn{4}{| c |}{\textbf{Cameras}}\\
	\hline
        \amazoncamera & Watch & \textbf{S1:} S-[627-634] C-[1229-1236] & 203 / 261 / 476\\
	\hline
        \blinkhub & Watch & \textbf{S1:} S-199 C-135 C-183 S-135 & 99 / 158 / 275\\
	\cline{2-4}
        & Photo & \textbf{S1:} S-199 C-135 C-183 S-135 & 87 / 173 / 774\\
	\hline
        \lefuncamera & Photo & \textbf{S1:} S-258 C-[206-210] S-386 C-206 & 17,871 / 19,032 / 20,358\\
        & & \textbf{S2:} C-222 S-198 C-434 S-446 C-462 S-194 C-1422 S-246 C-262 &\\
        & & \textbf{S3:} C-182 &\\
	\cline{2-4}
        & Recording & \textbf{S1:} S-258 C-210 S-386 C-206 & 13,209 / 15,279 / 16,302\\
        & & \textbf{S2:} C-222 S-198 C-434 S-446 C-462 S-194 &\\
	\cline{2-4}
        & Watch & \textbf{S1:} S-258 C-210 S-386 C-206 & 14,151 / 15,271 / 16,131\\
        & & \textbf{S2:} C-222 S-198 C-434 S-446 C-462 S-194 &\\
	\hline
        \microsevencamera & Watch & \textbf{S1:} D-242 PH-118 & 1 / 5 / 38\\
	\hline
        \zmododoorbell & Photo & \textbf{S1:} C-94 S-88 S-282 C-240 / \textbf{S1:} S-282 C-240 C-94 S-88 & 1,184 / 8,032 / 15,127\\
	\cline{2-4}
        & Recording & \textbf{S1:} C-94 S-88 S-282 C-240 / \textbf{S1:} S-282 C-240 C-94 S-88 & 305 / 7,739 / 15,137\\
	\cline{2-4}
        & Watch & \textbf{S1:} C-94 S-88 S-282 C-240 / \textbf{S1:} S-282 C-240 C-94 S-88 & 272 / 7,679 / 15,264\\
	\hline
		\multicolumn{4}{| c |}{\textbf{Light Bulbs}}\\
	\hline
        \flexbulb & ON/OFF & \textbf{S1:} PH-140 D-[346-347] & 4 / 44 / 78\\
	\cline{2-4}
        & Intensity & \textbf{S1:} PH-140 D-346 & 4 / 18 / 118\\
	\cline{2-4}
        & Color & \textbf{S1:} PH-140 D-346 & 4 / 12 / 113\\
	\hline
        \winkhub & ON/OFF & \textbf{S1:} PH-204 D-890 PH-188 D-113 & 43 / 55 / 195\\
	\cline{2-4}
        & Intensity & \textbf{S1:} PH-204 D-890 PH-188 D-113 & 43 / 50 / 70\\
	\cline{2-4}
        & Color & \textbf{S1:} PH-204 D-890 PH-188 D-113 & 43 / 55 / 106\\
	\hline
		\multicolumn{4}{| c |}{\textbf{Voice Command Devices}}\\
	\hline
        \allurespeaker & Audio ON/OFF & \textbf{S1:} C-658 C-412 & 89 / 152 / 196\\
	\cline{2-4}
        & Volume & \textbf{S1:} C-[594-602] & 217 / 4,010 / 11,005\\
				& & \textbf{S2:} C-[92-100] &\\
	\hline
        \echodot & Voice & \textbf{S1:} C-491 S-[148-179] & 1 / 23 / 61\\
	\cline{2-4}
        & Volume & \textbf{S1:} C-[283-290] C-[967-979] & 1,555 / 2,019 / 2,423\\
				& & \textbf{S2:} C-[197-200] C-[147-160] &\\
	\hline
        \echoplus & Audio ON/OFF & \textbf{S1:} S-100 C-100 & 1 / 5 / 28\\
	\cline{2-4}
        & Color & \textbf{S1:} S-100 C-100 & 1 / 4 / 18\\
	\cline{2-4}
        & Intensity & \textbf{S1:} S-100 C-100 & 1 / 4 / 11\\
	\cline{2-4}
        & Voice & \textbf{S1:} C-[761-767] S-437 & 1,417 / 1,871 / 2,084\\
				& & \textbf{S2:} C-172 S-434 &\\
	\cline{2-4}
        & Volume & \textbf{S1:} C-172 S-434 & 2 / 13 / 40\\
	\hline
        \echospot & Audio ON/OFF & \textbf{S1:} S-100 C-100 & 1 / 8 / 233\\
	\cline{2-4}
        & Voice & \textbf{S1:} C-246 S-214 & 1,220 / 1,465 / 1,813\\
				& & \textbf{S2:} C-172 S-434 &\\
	\cline{2-4}
        & Volume & \textbf{S1:} C-246 S-214 & 1,451 / 1,709 / 1,958\\
				& & \textbf{S2:} C-172 S-434 &\\
	\hline
        \googlehome & Voice & \textbf{S1:} C-1434 S-136 & 9 / 61 / 132\\
	\cline{2-4}
        & Volume & \textbf{S1:} C-1434 S-[124-151] & 8,020 / 9,732 / 10,002\\
				& & \textbf{S2:} C-521 S-[134-135] &\\
	\hline
        \googlehomemini & Voice & \textbf{S1:} C-1434 S-[127-153] & 1 / 29 / 112\\
	\cline{2-4}
        & Volume & \textbf{S1:} C-1434 S-[135-148] & 5 / 47 / 123\\
	\hline
        Harman Kardon & Voice & \textbf{S1:} S-1494 S-277 C-1494 & 2,199 / 2,651 / 3,762\\
				Invoke speaker& & \textbf{S2:} S-159 S-196 C-1494 &\\
	\cline{2-4}
        & Volume & \textbf{S1:} S-159 S-196 C-1418 C-1320 S-277 & 223 / 567 / 793\\
				& & \textbf{S2:} S-196 C-[404-406] &\\
	\hline
		\multicolumn{4}{| c |}{\textbf{Smart TVs}}\\
	\hline
        \firetv & Menu & \textbf{S1:} C-468 S-323 & 16 / 18 / 20\\
	\hline
        \lgtv & Menu & \textbf{S1:} PH-204 D-1368 PH-192 D-117 & 43 / 90 / 235\\
	\hline
        \rokutv & Remote & \textbf{S1:} PH-163 D-[163-165] & 578 / 1,000 / 1,262\\
        & & \textbf{S2:} PH-145 D-410 &\\
        & & \textbf{S2:} PH-147 D-113 &\\
	\hline
        \samsungtv & Menu & \textbf{S1:} PH-[237-242] D-274 & 2 / 7 / 15\\
	\hline
		\multicolumn{4}{| c |}{\textbf{Other Types of Devices}}\\
	\hline
        \honeywellthermostat & ON & \textbf{S1:} S-635 C-256 C-795 S-139 C-923 S-139 & 1,091 / 1,248 / 1,420\\
	\cline{2-3}
        & OFF & \textbf{S1:} S-651 C-256 C-795 S-139 C-923 S-139 & \\
	\cline{2-4}
        & Set & \textbf{S1:} C-779 S-139 & 86 / 102 / 132\\
	\hline
        \insteonhub & ON/OFF & \textbf{S1:} S-491 C-623 & 76 / 100 / 1,077\\
        & & \textbf{S2:} C-784 C-234 S-379 & \\
	\hline
        \samsungfridge & Set & \textbf{S1:} C-116 S-112 & 177 / 185 / 185\\
	\cline{2-4}
        & View Inside & \textbf{S1:} C-116 S-112 & 177 / 197 / 563\\
	\hline
  \end{tabular}
  }
  \end{center}
  \caption{Signatures extracted from the devices only in the \imcdataset{}~\cite{ren2019information} dataset.\label{tab:public-dataset-training}}
  }
\end{table*}

\begin{table*}[t!]
  \centering
  \majorrevision{
  \begin{center}
  { \footnotesize
  \begin{tabular}{| p{22mm} | p{12mm} | c | c | c | c | c | c |}
    \hline
        \textbf{Device} & \textbf{Event} & \textbf{Signature} & \textbf{Duration (ms)} & \multicolumn{4}{ c |}{\textbf{Matching}}\\
    \cline{5-8}
        & & & \textbf{Min./Avg./Max./St.Dev.} & \textbf{WAN} & \textbf{FPR} & \textbf{Wi-Fi} & \textbf{FPR}\\
        & & & & \textbf{Sniffer} & & \textbf{Sniffer} & \\
	\hline
        \wemoinsightplug & ON/OFF & $\ast$\textbf{S1:} PH-475 D-246 & 29 / 33 / 112 / 9 & - & - & 98.75\% & 0\\
    \cline{3-4}
        & & $\dagger$\textbf{S1:} PH-475 D-246 & 31 / 42 / 111 / 15 & & & & \\
	\hline
        \blinkcamera & Watch & $\ast$\textbf{S1:} C-331 S-299 C-139 & 267 / 273 / 331 / 8 & 100\% & 0 & 100\% & 0\\
	\cline{3-4}
        & & $\dagger$\textbf{S1:} C-331 S-299 C-139 & 170 / 269 / 289 / 19 & & & & \\
  	\cline{2-8}
        & Photo & $\ast$\textbf{S1:} C-331 C-123 S-139 S-123 S-187 C-1467 & 281 / 644 / 1,299 / 348 & 97.37\% & 0 & 97.50\% & 0\\
	\cline{3-4}
        & & $\dagger$\textbf{S1:} C-331 C-123 S-139 S-123 S-187 C-1467 & 281 / 742 / 2,493 / 745 & & & & \\
	\hline
        \tplinkplug & ON & $\ast$\textbf{S1:} C-592 S-1234 S-100 & 70 / 74 / 85 / 2 & 100\% & 0 & - & - \\
    \cline{2-3}
        (\devicecloudcomm) & OFF & $\ast$\textbf{S1:} C-593 S-1235 S-100 & & & & & \\
	\cline{2-4}
        & ON & $\dagger$\textbf{S1:} C-605 S-1213 S-100 & 16 / 19 / 29 / 2 & & & & \\
    \cline{2-3}
        & OFF & $\dagger$\textbf{S1:} C-606 S-1214 S-100 & & & & & \\
	\hline
        \tplinkplug & ON & $\ast$\textbf{S1:} PH-172 D-115 & 406 / 743 / 10,667 / 1,417 & - & - & 100\% & 0\\
        (\phonedevicecomm \& &  & \textbf{S2:} C-592 S-1234 S-100 & & & & & \\
	\cline{2-3}
        \devicecloudcomm) & OFF & $\ast$\textbf{S1:} PH-172 D-115 & & & & & \\
        &  & \textbf{S2:} C-593 S-1235 S-100 & & & & & \\
	\cline{2-4}
        & ON & $\dagger$\textbf{S1:} PH-172 D-115 & 197 / 382 / 663 / 165 & & & & \\
        &  & \textbf{S2:} C-605 S-1213 S-100 & & & & & \\
	\cline{2-3}
        & OFF & $\dagger$\textbf{S1:} PH-172 D-115 & & & & & \\
        &  & \textbf{S2:} C-606 S-1214 S-100 & & & & & \\
	\hline
        \sengledbulb & ON & $\ast$\textbf{S1:} S-[217-218] C-[209-210] & 4,304 / 6,238 / 8,145 / 886 & - & - & - & - \\
        & & \textbf{S2:} C-430 & & & & & \\
        & & \textbf{S3:} C-466 & & & & & \\
    \cline{2-3}
        & OFF & $\ast$\textbf{S1:} S-[217-218] C-[209-210] & & & & & \\
        & & \textbf{S2:} C-430 & & & & & \\
        & & \textbf{S3:} C-465 & & & & & \\
	\cline{2-4}
        & ON & $\dagger$\textbf{S1:} S-219 C-210 & 354 / 2,590 / 3,836 / 859 & & & & \\
        & & \textbf{S2:} C-428 & & & & & \\
        & & \textbf{S3:} C-[478-479] & & & & & \\
    \cline{2-3}
        & OFF & $\dagger$\textbf{S1:} S-219 C-210 & & & & & \\
        & & \textbf{S2:} C-428 & & & & & \\
        & & \textbf{S3:} C-[478-480] & & & & & \\
	\hline
        \tplinkbulb & ON & $\ast$\textbf{S1:} PH-258 D-288 & 8 / 77 / 148 / 42 & - & - & - & - \\
    \cline{2-3}
        & OFF & $\ast$\textbf{S1:} PH-258 D-305 & & & & & \\
	\cline{2-4}
        & ON & $\dagger$\textbf{S1:} PH-258 D-227 & 17 / 92 / 224 / 46 & & & & \\
    \cline{2-3}
        & OFF & $\dagger$\textbf{S1:} PH-258 D-244 & & & & & \\
	\hline
  \end{tabular}
  }
  \end{center}
  \caption{Common devices in  the \imcdataset{} and our testbed experiments. 
	$\ast$ signature: training on our testbed.
	$\dagger$ signature: training on \imcdataset{}~\cite{ren2019information}. 
	Matching: training on testbed, detection on \imcdataset{}.
	The number of events vary (around 30-40) per \event type---the result is presented in  \% for convenience.
	\label{tab:public-dataset-training-detection}}
  }
	\vspace{-1em}
\end{table*}

\label{sect:public-dataset}
In this section, we apply the \tool{} methodology  to a state-of-the-art, publicly available IoT dataset: the \imcdataset{} dataset~\cite{ren2019information}. First, we show that \tool{}  successfully extracted signatures from new devices in this dataset, thus validating the generality of the methodology and expanding our coverage of devices. Then,  we compare the signatures extracted from the \imcdataset{} dataset to those extracted from our testbed dataset, for devices that were present in both.

\myparagraph{The \imcdataset{} Dataset.}
The \imcdataset{} dataset~\cite{ren2019information} contains network traces from 55 distinct IoT devices.\footnote{\majorrevision{The paper \cite{ren2019information} reports results from 81 physical devices, but 26 device models are present in both the US and the UK testbed, thus there are only 55 distinct models.}}
Each PCAP file in the dataset contains traffic observed for a single device during a short timeframe surrounding a single \event{} on that device.
Moreover, the authors provide timestamps for when they performed each \event{}.
As a result, we can merge all PCAP files for each device and \event type combination into a single PCAP file, and directly apply \tool{} to extract signatures, similarly to how we extracted signatures from the training set we collected using our testbed.  

We only considered a subset of the 55 devices in the \imcdataset{} dataset, due to a combination of limitations of the dataset and of our methodology. In particular, we did not apply \tool{} to the following groups of devices in the \imcdataset{} dataset:
(1) 3 devices with nearly all PCAP files empty;
(2) 6 devices  with a limited number (three or less) of \event samples;\footnote{We consider this to be too few samples to have confidence in the extracted signatures. In contrast, the traces for the remaining devices generally had 30--40 \event samples for each device and \event type combination.} and
(3) 13 devices that only communicate via UDP (\tool{}'s current  implementation  only considers TCP traffic).
Next, we report results from applying \tool{} to the remaining 33 devices in the \imcdataset{} dataset. Out of those, 26 are exclusive to the \imcdataset{} dataset, while seven are common across the \imcdataset{} dataset and our testbed dataset.

\myparagraph{Devices only in the \imcdataset{} Dataset.}
We ran \tool{}'s signature extraction on the traces from the 26 new devices from the \imcdataset{} dataset. 
\tool{} successfully extracted signatures for 21  devices and we summarize those signatures in  Table~\ref{tab:public-dataset-training}.\techreport{\footnote{For some of the devices, we had to relax the training time window, \ie $t=30\textrm{s}$, because there is usually a gap of more than 20s from the provided timestamps to the appearance of event-related traffic. As a result, \tool extracted longer signatures for certain devices, \eg the \lefuncamera. Further, we also had to exclude some PCAP files that are empty or contain traffic that look inconsistent with the majority (\eg some only contain DNS or IGMP traffic).}}  Some of these devices provide similar functionality as those in our testbed dataset (\eg bulbs, cameras). Interestingly, we were also able to successfully extract signatures for many new types of devices that we did not have access to during our testbed experiments. Examples include voice-activated devices, smart TVs, and even a fridge. This validates the generality of the \tool{} methodology and greatly expands our coverage of devices.

There were also 5, out of 26, new devices that \tool{} originally appeared to not extract signatures from. However, upon closer inspection of their PCAP files and \tool{}'s output logs,  we observed that those devices did actually exhibit a new type of signature that we had not previously encountered in our testbed experiments:  a sequence of packet pairs with the exact same pair of packet lengths for the same event.  
The default configuration of \tool{} would have discarded the clusters of these  packet pairs during the signature creation of the training phase (see Section~\ref{sect:system-training}), because the number of occurrences of these pairs is higher than (in fact a multiple of) the number of events. However, based on this intuitive observation,  \tool{} can easily be adapted to  extract those signatures as well: it can take into account the timing of packet pairs in the same cluster instead of only across different clusters, and concatenate them into longer sequences. We note that these frequent packet pairs can be either new signatures for new devices, or can be due to periodic background communication.  Unfortunately, the \imcdataset{} dataset does not include network traces for idle periods (where no events are generated), thus we cannot confirm or reject this hypothesis.

\myparagraph{Common Devices.}
We next report our findings for devices that are present in both the \imcdataset{} dataset and in our own testbed dataset, referred to as common devices.
There were already 6 common devices across the 2 datasets, and we acquired an additional device after consulting with the authors of the paper: the \blinkcamera.
We excluded 2 common devices: (1) the \nestthermostat as it was tested for different \event types; and (2) the \huebulb as it has a unique signature that \tool cannot use to perform 
matching---it is a combination of \devicecloudcomm (visible only to the \wansniffer)
and \phonedevicecomm communications (visible only to the \wifisniffer). 
Table~\ref{tab:public-dataset-training-detection} summarizes the results for the 5 remaining common 
devices. 
%
 First, we report the complete signatures extracted from each dataset. 
The signatures reported in Table~\ref{tab:summary-signatures} were obtained from data collected throughout 2018. For the \wemoinsightplug and \tplinkplug, we repeated our testbed data collection and signature extraction in December 2019 to facilitate a better comparison of signatures from the same devices across different points in time. 
Then, we compare the signatures extracted from the two datasets for the common devices: some of the signatures are identical and some are similar. Such a comparison provides more information than simply training on one dataset and testing on the other. 

\vspace{+1em}
\mysubparagraph{Identical Signatures.}
For the \wemoinsightplug and \blinkcamera, the signatures extracted from the \imcdataset{} dataset and our dataset (December 2019) were identical.
Since the signatures obtained from our own dataset do not have any variations in packet lengths, we used \tool{}'s exact matching strategy (see Section~\ref{sect:system-detection}) to detect events in the \imcdataset{} dataset, and we observed a recall rate of 97\% or higher for both devices (see Table~\ref{tab:public-dataset-training-detection}).

\mysubparagraph{Similar Signatures.}
For the \tplinkplug and \sengledbulb, the signatures extracted from the \imcdataset{} dataset are slightly different from those extracted from our own dataset: some packet lengths at certain positions in the sequence are different (by a few and up to tens of bytes), and these differences appear to be consistent (\ie{} all signatures from both datasets appear to be completely deterministic as they do not contain packet length ranges). For example, the \tplinkplug's ON \event is  
\code{C-592 S-1234 S-100} in our experiment vs. \code{C-605 S-1213 S-100} in \imcdataset{}.
To understand the cause of these discrepancies, we examined the \tplinkplug in further detail---between the two devices, its signatures exhibit the largest difference in packet lengths across datasets.
Through additional experiments on the \tplinkplug, we identified that changes to configuration parameters (\eg user credentials of different lengths) could cause the packet lengths to change. However, the packet lengths are deterministic for each particular set of user credentials. 

For devices that exhibit this kind of behavior, an attacker must first train \tool{} multiple times with different user credentials to determine to what extent these configuration changes affect the packet lengths in the signatures.
Moreover, the signature matching strategy should not be exact, but must be relaxed to allow for small variations in packet lengths.
To this end, we implemented \emph{relaxed matching} that augments the matching strategies discussed in Section~\ref{sect:system-detection}.\footnote{In relaxed matching, a delta equal to the greatest variation observed in packet lengths is applied to 
the packets that vary due to configuration changes.
For the \tplinkplug, we observed that the first packets differ by 13B in the
the \devicecloudcomm signatures from the two datasets (\ie $13=605-592=606-593$)
and the second packets differ by 21B (\ie $21=1234-1213=1235-1214$), thus a delta of 21B is used.}
We ran \tool with relaxed matching on the \tplinkplug with a delta of 21B, and 
successfully detected 100\% of events.
Furthermore, by performing the negative control experiments described in Section~\ref{sect:negative-control}, we verified that the increase in FPR due to 
relaxed matching is negiligible.
For dataset 1, relaxed matching results in two FPs for the \wifisniffer{}.
For dataset 3, relaxed matching results in seven FPs for the \wifisniffer{} and one FP for the \wansniffer{}. 
In comparison, exact matching only produces one false positive for the \wifisniffer{} for dataset 3.
We note that the total number of packets across these datasets is 440 million.
However, relaxed matching may eliminate the ability to distinguish event types for signatures that only differ by a few bytes (\eg the packet lengths for the \tplinkplug{}'s ON and OFF signatures differ by one byte).

\mysubparagraph{Signature Evolution.}
We observed that some signatures change over time, presumably due to changes to the device's communication protocol.
The \wemoinsightplug{}'s signature changed slightly from our earlier dataset from 2018 (see Table~\ref{tab:summary-signatures}) to our latest dataset collected in December 2019 (see Table~\ref{tab:public-dataset-training-detection}): the first PH-259 packet is no longer part of the signature. 
Both of these datasets were collected using the same testbed with the same user accounts, but with different device firmware versions.
Therefore, the change is probably due to changes in the communication protocol, introduced in firmware updates.
This is further backed by the observation that the \wemoinsightplug's signature extracted from the \imcdataset{} dataset
(collected in April 2019) is identical to the signature extracted from our December 2019's dataset.
This implies that there has been a protocol change between 2018 and April 2019, but the protocol has then remained unchanged until December 2019.

Similarly, the \tplinkbulb{}'s signature has changed slightly from our first to our second in-house dataset (see Tables~\ref{tab:summary-signatures} and~\ref{tab:public-dataset-training-detection}), and is also slightly different for the \imcdataset{} dataset.\footnote{We also repeated our experiments for the \tplinkbulb to further understand this phenomenon.}
The signatures from our 2018 dataset and those from the \imcdataset{} dataset differ in the first packet (\code{PH-198} vs. \code{PH-258}, an offset of 60 bytes), and the signatures from the \imcdataset{} dataset and those from our December 2019 differ in the second packet (\code{D-227} vs. \code{D-288} and \code{D-244} vs. \code{305}, an offset of 61 bytes). 
Thus, we also suspect that there is a signature evolution due to firmware updates for the \tplinkbulb.
Signature evolution is a limitation of our approach, and is elaborated on in Section~\ref{sect:conclusion}.
Nevertheless, an attacker can easily overcome this limitation simply by repeating
\tool's training to extract the latest signatures from a device right before launching an attack.
}

\rahmedit{
\subsection{Parameters Selection and Sensitivity}
\label{sect:parameter-sensitivity}
\myparagraph{Clustering Parameters.}
We empirically examined a range of values for the parameters
of the DBSCAN algorithm. 
We tried all combinations 
of $\epsilon \in \left\{ 1, 2,..., 10 \right\}$ and
$\mathtt{minPts} \in \left\{ 30, 31, ..., 50 \right\}$.
For those devices that exhibit no variation in their signature related 
packet lengths, \eg the \tplinkplug, the output of the clustering remains stable
for all values of $\epsilon$ and $\mathtt{minPts} < 50$. 
For such devices, keeping $\epsilon$ at a minimum and $\mathtt{minPts}$ close 
to the number of events $n$ reduces the number of noise points that become part of the resulting clusters.
However, our experiments  show that there is a tradeoff in
applying strict
bounds to devices with more variation in their packet lengths
(\eg the \dlinkplug), strict bounds can result in  losing
clusters that contain packet pairs related to \events. For the \dlinkplug, this happens if 
$\epsilon < 7$ and $\mathtt{minPts} > 47$. 
In our experiments, we used our initial values of $\epsilon=10$ and $\mathtt{minPts}=45$
(\ie $\mathtt{minPts}=\lfloor n-0.1n \rfloor$ with $n$ = number of expected events)
from our smart plugs experiment (\ie the \tplinkplug, \dlinkplug,
and \smartthingsplug) that allowed \tool to produce the packet-level signatures
we initially observed manually (see Section~\ref{sect:network-behavior}).
We then used them as default parameters for \tool to analyze new devices and
extracted packet-level signatures from 15 more devices.

\myparagraph{Time Window and Signature Duration.}
We also measured the duration of our signatures---defined as the time
 between the first and the last packets of the signature.
\majorrevision{
Table~\ref{tab:summary-signatures} reports all the results.}
The longest signature duration measured is 9,132 ms (less than 10 seconds) for
the \sengledbulb's ON/OFF signatures from the \phonecloudcomm communication. 
This justifies our choice of training time window \timewindow = 15 seconds
during trace filtering and signature validation 
(see Section~\ref{sect:system-training}).  This conservative  choice 
also provides slack to accommodate other devices that we have not evaluated and that may have 
signatures with a longer duration.
This implies that \events can be generated every 15 seconds or longer.
We conservatively chose this duration to be 131 seconds to give ample time for
an \event to finish, and to easily separate false positives from true positives.

\section{Possible Defenses}
\label{sect:defenses}
There are several broad approaches that can obfuscate network traffic to defend against passive inference attacks that analyze network traffic metadata:
\squishcount
\item {\em Packet padding} adds dummy bytes to each packet to confuse inference techniques that rely on individual packet lengths, and less so volume. Packets can be padded to a fixed length (\eg{} MTU) or with a random number of bytes. 
\item {\em Traffic shaping} purposely delays packets to confuse inference techniques that rely on packet inter-arrival times and volume over time. 
\item  {\em Traffic injection} adds dummy packets in patterns that look similar (\eg have the same lengths, inter-arrival times or volume signature \etc) as the real events, thus  hiding the real \event{} traffic in a crowd of fake \events{}. 
\vspace{-5pt}
\countend
The above approaches can be implemented in different ways and can also be combined (\eg on the same VPN).
Since our signatures rely on unique sequences of individual packet lengths, packet padding is the most natural defense and therefore discussed in depth below.
We defer discussion of traffic shaping and traffic injection to \appdx{Appendix~\ref{sect:defenses-packet-injection-cr}}.
We first provide a brief overview of packet padding in the literature.
We then discuss how packet padding may be implemented to obfuscate packet-level signatures. 
Finally, we evaluate the efficacy of packet padding for the \tplinkplug{}.

%
%
	
\myparagraph{Packet Padding in the Literature.}
Packet padding has already been 
studied as a countermeasure for website 
fingerprinting~\cite{liberatore2006inferring, dyer2012peek, cai2014cs, cai2014systematic}.
Liberatore and Levine~\cite{liberatore2006inferring} showed that padding to MTU drastically reduces the accuracy of a Jaccard coefficient based classifier and a naive Bayes classifier, both of which use a feature set very similar to packet-level signatures: a vector of $<$direction, packet length$>$ tuples.
Dyer \etal{}~\cite{dyer2012peek} later showed that such padding is less successful against more advanced classifiers, such as the support vector machine proposed by Panchenko \etal{}~\cite{panchenko2011website} that also considers coarse-grained features such as total traffic volume.
Cai \etal{}~\cite{cai2014cs,cai2014systematic} improved~\cite{dyer2012peek} by providing a strategy to control traffic flow that better obfuscates the traffic volume as a result.
	%
Although applied in a different context, these prior works indicate that packet padding should successfully guard against a packet-level signature attack.
The question then becomes where and how to implement the padding mechanism.
	
\myparagraph{Possible Implementations.}
Next, we discuss the potential benefits and drawbacks of different packet padding implementations.
We consider a VPN-based implementation, padding at the application layer, and TLS-based padding. 
	
\mysubparagraph{VPN.}
One option is to route traffic from the smart home devices and the smartphone through a VPN that  
pads outbound tunneled packets with dummy bytes and strips the padding off of inbound tunneled packets: a technique also considered in~\cite{princeton-stp}.
The smart home end of the VPN may be implemented either directly on each  device and smartphone or on a middlebox, \eg{} the home router.
The former provides protection against \emph{both} the WAN and Wi-Fi sniffers as the padding is preserved on the local wireless link, whereas the latter only defends against the \wansniffer{}.
However, an on-device VPN may be impractical on devices with limited software stacks and/or computational resources.
The middlebox-based approach may be used to patch existing devices without changes to their software.
Pinheiro et al.~\cite{pinheiro2018} provide an implementation in which the router is the client-side end of  the VPN, and where the padding is added to the Ethernet trailer.

	
\mysubparagraph{Application Layer and TLS.}
Another option is to perform the padding at the application layer.
This has at least three benefits: (1)~it preserves the padding across all links, thus provides protection against both the WAN and Wi-Fi sniffers; (2)~it imposes no work on the end user to configure their router to use a VPN; and (3)~it can be implemented entirely in software. 
An example is HTTPOS by Luo et al.~\cite{luo2011}, which randomizes the lengths of HTTP requests (\eg{} by adding superfluous data to the HTTP header).
%
One drawback of application layer padding is that it imposes extra work on the application developer.
This may be addressed  by including the padding mechanism in libraries for standardized protocols (\eg{} OkHttp~\cite{okhttp}), but a separate implementation is still required for every proprietary protocol.
A better alternative is to add the padding between the network and application layers.
This preserves the benefits of application layer padding highlighted above, but eliminates the need for the application developer to handle padding.
As suggested in~\cite{dyer2012peek}, one can use the padding functionality that is already available in TLS~\cite{rfctls13}.

\myparagraph{Residual Side-Channel Information.}
Even after  packet padding is applied, there may still be other  side-channels,  \eg timing and packet directions, and/or coarse-grained features such as total volume,
total number of packets, and burstiness, as demonstrated by~\cite{dyer2012peek}.
Fortunately, timing information (\eg{} packet inter-arrival times and duration of the entire packet exchange) is highly location dependent (see the comparison of signature durations in Table~\ref{tab:public-dataset-training-detection}), as it is impacted by the propagation delay between the home and the cloud, as well as the queuing and transmission delays on individual links on this path. 
Exploiting timing information requires a much more powerful adversary: one that uses data obtained from a location close to the smart home under attack.
The work of Apthorpe \etal{} on traffic shaping~\cite{princeton-spying} and stochastic traffic padding (STP)~\cite{princeton-stp} 
may aid in obfuscating timing, volume, and burstiness.

\myparagraph{Efficacy of Packet Padding.} The discussion has been qualitative so far. Next, we perform a simple test to empirically assess the efficacy of packet padding for the \tplinkplug{}.
	
\mysubparagraph{Setup.}
We simulated padding to the MTU by post-processing the \tplinkplug{}  testbed trace from Section~\ref{sect:detection-phase} (50 ON and 50 OFF \events, mixed with background traffic) using a simplified version of \tool{}'s detection that only considers the order and directions of packets, but pays no attention to the packet lengths.
We focus on the \wansniffer{} because it is the most powerful adversary: it can separate traffic into individual TCP connections and eliminate the confusion that arises from multiplexing.
We used the \tplinkplug's two-packet signatures for ON and OFF \events (see Table~\ref{tab:summary-signatures}) as the \phonedevicecomm communication is not visible on the WAN.
We consider the packet padding's impact on transmission and processing delays to be negligible.
We assume that the adversary uses \emph{all} available information to filter out irrelevant traffic.
Specifically, the adversary uses the timing information observed during training to only consider request-reply exchanges that comply with the signature duration.\footnote{$t=0.204\textrm{s} \implies \lceil 0.204 + 0.1 \times 0.204\textrm{s} \rceil = 0.224\textrm{s}$ (see Table~\ref{tab:summary-signatures} and \appdx{Appendix~\ref{sect:system-detection-cr}})}
Moreover, since the \tplinkplug{} uses TLSv1.2 (which does not encrypt the SNI), the adversary can filter the trace to only consider TLS Application Data packets to the relevant TP-Link host(s) in the no-VPN scenarios.
	
\mysubparagraph{VPN-Based Padding.}
%
To simulate VPN-based packet padding, we consider all packets in the trace as multiplexed over a single connection and perform signature detection on this tunnel.
This results in a total of 193,338 positives, or, put differently, more than 1,900 false positives for every \event{}.
This demonstrates that VPN-based packet padding works well for devices with short signatures (\eg a single packet pair).

\mysubparagraph{TLS-Based Padding.}
From the training data, the adversary knows that the signature is present in the \tplinkplug{}'s communication with \code{events.tplinkra.com}.
To simulate TLS-based packet padding, we performed signature detection on the TLS Application Data packets of each individual TLSv1.2 connection with said host.
As expected, this produced a total of 100 detected events, with no FPs. 
Intuitively, this is because the only TLS Application Data packets of these connections are exactly the two signature packets, and the device only communicates with this domain when an event occurs.
		
\mysubparagraph{Hybrid.}
We next explore how multiplexing all of the \tplinkplug{}'s traffic over a single connection affects the false positives (the plug communicates with other TP-Link hosts).\techreport{\footnote{We envision that this could be implemented by maintaining a single TLS connection between the device and a single TP-Link endpoint, $T$, that would then carry all application layer messages, each prepended with an additional header that identifies the type and order of that particular request/response, and padded to MTU using TLS record padding.		For each request, $T$ would interpret its type based on the application layer header, and forward it to an appropriate backing server responsible for that part of the application logic (\ie $T$ is analogous to a load balancer).}}
This is conceptually similar to a VPN, but only tunnels application layer protocols and can be implemented in user space (without TUN/TAP support).
To simulate such a setup, we filtered the trace to only contain IPv4 unicast traffic to/from the \tplinkplug{}, 
and dropped all packets that were not TLS Application Data.
We then performed detection on the TLS Application Data packets, treating them as belonging to a single TLS connection.
For this scenario, we observed 171 positives. 
While significantly better than TLS-based padding for individual TLS connections, the attacker still has a high probability (more than 50\%) of guessing the occurrence of each event (but cannot distinguish ON from OFF). 

\mysubparagraph{Recommendations.}
Based on the above insights, we recommend VPN-based packet padding due to its additional obfuscation (encryption of the Internet endpoint and multiplexing of IoT traffic with other traffic) as TLS-based padding seems insufficient for devices with simple signatures and little background traffic.
For more chatty devices, multiplexing all device traffic over a single TLS connection to a single server may provide sufficient obfuscation at little overhead.

}

\section{Conclusion and \majorrevision{Discussion}}
\label{sect:conclusion}

\myparagraph{Summary.}
We designed, implemented, and evaluated \tool, a  methodology for
automatically extracting packet-level signatures for smart home
device events from network traffic. Notably, traffic can be encrypted or generated by proprietary or unknown protocols.  This work advances the
state-of-the-art by: (1) identifying simple packet-level signatures that were not previously known; (2) proposing an automated methodology for extracting these signatures from training datasets; and (3) showing that they are effective in inferring events across a wide range of devices, \event types, traces, and attack models  (\wansniffer and \wifisniffer).  
We have made \tool (software and datasets) publicly available at~\cite{pingpong-software}.
We note that the new packet-level signatures can be used for several applications, including launching a passive inference attack, anomaly detection, \etc{}
To deal with such attacks, we outlined a simple defense based on packet padding.

\majorrevision
{
\myparagraph{Current Limitations and Future Directions.}
\tool{} has its limitations and can be extended  in several directions.
	
	First, in order to extract the signature of a new device, one must first acquire the device and apply \tool{} to train and extract the corresponding packet-level signatures. This is actually realistic for an attacker with minimal side information, \ie one who knows what device they want to attack or who wants to distinguish between two different types of devices. 
One direction for future work is to extend \tool{} by finding ``similar'' known behaviors for a new device, \eg via relaxed matching of known and unknown signatures.

	Second, a signature may evolve over time, \eg when a software/firmware update changes a device's communication protocol. Whoever maintains the signature  (\eg the attacker) needs to retrain and update the signature. 
	We observed this phenomenon, for example, for the \tplinkplug{}. 
	This  can be handled by relaxed matching since the packet sequences tend to be mostly stable and only evolve by a few bytes (see Section~\ref{sect:public-dataset}). 

	Third,  there may be inherent variability in some signatures 
	due to configuration parameters (\eg credentials and device IDs) that are sent to the cloud and may lead to slightly different packet lengths. In Section~\ref{sect:public-dataset}, we saw that this variability is small: from a few to tens of bytes difference and only for some individual packets within a  longer sequence. An attacker could train with different configuration parameters and apply relaxed matching when necessary (only on packets with length variations). 
	
	Other possible improvements include: profiling and subtracting background/periodic traffic during signature creation, and unifying the way we account for small variation in the signatures in the training 
	and detection---\tool currently supports range-based matching (\appdx{see Appendix~\ref{sect:system-detection-cr}}) and relaxed matching as separate features. 
	Another limitation is that our methodology currently applies only to TCP---not to UDP-based devices that do not follow the request-reply pattern.

	\myparagraph{Conclusion.} We believe that the new packet-level signatures identified by \tool{} are a simple, intuitive, and universal means for profiling IoT devices. However, we see \tool{} only as a building block, which is part of a bigger toolbox for IoT network traffic analysis. We believe that it can and should be combined with other  complementary ideas, \eg traffic shape/volume-based signatures, semi-supervised learning, \etc{}
}

\section*{Acknowledgment}
\label{sect:acknowledgment}

This project was supported by the National Science
Foundation under grants CNS-1649372, 
CNS-1703598, OAC-1740210, CNS-1815666, CNS-1900654 and  a UCI Seed Funding Award at UCI.
The authors would like to thank the anonymous NDSS reviewers for their valuable feedback,
which helped significantly improve the paper. 
We would also like to thank Anastasia Shuba for her insights and advice during the project's early stages.

{\footnotesize
\bibliographystyle{abbrv}
\bibliography{references}

\begin{thebibliography}{10}

\bibitem{ifttt}
{IFTTT}.
\newblock \url{https://www.ifttt.com/}, September 2018.

\bibitem{ifttt-motion}
If motion detected by {D-Link} motion sensor, then turn on {D-Link} smart plug.
\newblock
  \url{https://ifttt.com/applets/393508p-if-motion-detected-by-d-link-motion-sensor-then-turn-on-d-link-smart-plug},
  January 2020.

\bibitem{acar2018peek}
A.~Acar, H.~Fereidooni, T.~Abera, A.~K. Sikder, M.~Miettinen, H.~Aksu,
  M.~Conti, A.-R. Sadeghi, and A.~S. Uluagac.
\newblock {Peek-a-Boo}: I see your smart home activities, even encrypted!
\newblock {\em arXiv preprint arXiv:1808.02741}, 2018.

\bibitem{topwebsites}
Alexa.
\newblock Top sites in {United States}.
\newblock \url{https://www.alexa.com/topsites/countries/US}, November 2018.

\bibitem{sok-alrawi}
O.~Alrawi, C.~Lever, M.~Antonakakis, and F.~Monrose.
\newblock {SoK}: Security evaluation of home-based {IoT} deployments.
\newblock In {\em 2019 2019 IEEE Symposium on Security and Privacy (SP)},
  volume~00, pages 208--226.

\bibitem{yourthings}
O.~Alrawi, C.~Lever, M.~Antonakakis, and F.~Monrose.
\newblock Yourthings scorecard.
\newblock \url{https://yourthings.info/}, 2019.

\bibitem{topdevices}
Amazon.
\newblock
  \url{https://www.amazon.com/smart-home/b/?ie=UTF8&node=6563140011&ref_=sv_hg_7},
  March 2019.

\bibitem{anderson-identifying-encrypted-malware}
B.~Anderson and D.~McGrew.
\newblock Identifying encrypted malware traffic with contextual flow data.
\newblock In {\em Proceedings of the 2016 ACM Workshop on Artificial
  Intelligence and Security}, AISec '16, pages 35--46, New York, NY, USA, 2016.
  ACM.

\bibitem{adbarticle}
Android.com.
\newblock Android debug bridge (adb).
\newblock \url{https://developer.android.com/studio/command-line/adb}, November
  2018.

\bibitem{princeton-stp}
N.~Apthorpe, D.~Y. Huang, D.~Reisman, A.~Narayanan, and N.~Feamster.
\newblock Keeping the smart home private with smart(er) {IoT} traffic shaping.
\newblock {\em Proceedings on Privacy Enhancing Technologies}, 2019(3), 2019.

\bibitem{princeton-spying-blinds}
N.~Apthorpe, D.~Reisman, and N.~Feamster.
\newblock Closing the blinds: Four strategies for protecting smart home privacy
  from network observers.
\newblock {\em CoRR}, abs/1705.06809, 2017.

\bibitem{princeton-spying-castle}
N.~Apthorpe, D.~Reisman, and N.~Feamster.
\newblock A smart home is no castle: Privacy vulnerabilities of encrypted {IoT}
  traffic.
\newblock {\em CoRR}, abs/1705.06805, 2017.

\bibitem{princeton-spying}
N.~Apthorpe, D.~Reisman, S.~Sundaresan, A.~Narayanan, and N.~Feamster.
\newblock Spying on the smart home: Privacy attacks and defenses on encrypted
  {IoT} traffic.
\newblock {\em CoRR}, abs/1708.05044, 2017.

\bibitem{bissiaspets}
G.~D. Bissias, M.~Liberatore, D.~Jensen, and B.~N. Levine.
\newblock Privacy vulnerabilities in encrypted {HTTP} streams.
\newblock In {\em Proceedings of the 5th International Conference on Privacy
  Enhancing Technologies}, PET'05, pages 1--11, Berlin, Heidelberg, 2006.
  Springer-Verlag.

\bibitem{cai2014cs}
X.~Cai, R.~Nithyanand, and R.~Johnson.
\newblock {CS-BuFLO}: A congestion sensitive website fingerprinting defense.
\newblock In {\em Proceedings of the 13th Workshop on Privacy in the Electronic
  Society}, pages 121--130. ACM, 2014.

\bibitem{cai2014systematic}
X.~Cai, R.~Nithyanand, T.~Wang, R.~Johnson, and I.~Goldberg.
\newblock A systematic approach to developing and evaluating website
  fingerprinting defenses.
\newblock In {\em Proceedings of the 2014 ACM SIGSAC Conference on Computer and
  Communications Security}, pages 227--238. ACM, 2014.

\bibitem{cai2012}
X.~Cai, X.~C. Zhang, B.~Joshi, and R.~Johnson.
\newblock Touching from a distance: Website fingerprinting attacks and
  defenses.
\newblock In {\em Proceedings of the 2012 ACM Conference on Computer and
  Communications Security}, CCS '12, pages 605--616, New York, NY, USA, 2012.
  ACM.

\bibitem{chen2010side}
S.~Chen, R.~Wang, X.~Wang, and K.~Zhang.
\newblock Side-channel leaks in web applications: A reality today, a challenge
  tomorrow.
\newblock In {\em 2010 IEEE Symposium on Security and Privacy}, pages 191--206.
  IEEE, 2010.

\bibitem{copos2016anybody}
B.~Copos, K.~Levitt, M.~Bishop, and J.~Rowe.
\newblock Is anybody home? {Inferring} activity from smart home network
  traffic.
\newblock In {\em Security and Privacy Workshops (SPW), 2016 IEEE}, pages
  245--251. IEEE, 2016.

\bibitem{ml-ddos}
R.~Doshi, N.~Apthorpe, and N.~Feamster.
\newblock Machine learning {DDoS} detection for consumer internet of things
  devices.
\newblock {\em CoRR}, abs/1804.04159, 2018.

\bibitem{dyer2012peek}
K.~P. Dyer, S.~E. Coull, T.~Ristenpart, and T.~Shrimpton.
\newblock Peek-a-boo, i still see you: Why efficient traffic analysis
  countermeasures fail.
\newblock In {\em 2012 IEEE symposium on security and privacy}, pages 332--346.
  IEEE, 2012.

\bibitem{dbscan}
M.~Ester, H.-P. Kriegel, J.~Sander, X.~Xu, et~al.
\newblock A density-based algorithm for discovering clusters in large spatial
  databases with noise.
\newblock In {\em Kdd}, volume~96, pages 226--231, 1996.

\bibitem{ghiglieri2014}
M.~Ghiglieri and E.~Tews.
\newblock {A privacy protection system for HbbTV in Smart TVs}.
\newblock In {\em 2014 IEEE 11th Consumer Communications and Networking
  Conference (CCNC)}, pages 357--362, Jan 2014.

\bibitem{hayes2016}
J.~Hayes and G.~Danezis.
\newblock K-fingerprinting: A robust scalable website fingerprinting technique.
\newblock In {\em Proceedings of the 25th USENIX Conference on Security
  Symposium}, SEC'16, pages 1187--1203, Berkeley, CA, USA, 2016. USENIX
  Association.

\bibitem{herrmann2009website}
D.~Herrmann, R.~Wendolsky, and H.~Federrath.
\newblock Website fingerprinting: {Attacking} popular privacy enhancing
  technologies with the multinomial na{\"\i}ve-bayes classifier.
\newblock In {\em Proceedings of the 2009 ACM workshop on Cloud computing
  security}, pages 31--42. ACM, 2009.

\bibitem{jin-unveiling}
Y.~Jin, E.~Sharafuddin, and Z.-L. Zhang.
\newblock Unveiling core network-wide communication patterns through
  application traffic activity graph decomposition.
\newblock In {\em Proceedings of the Eleventh International Joint Conference on
  Measurement and Modeling of Computer Systems}, SIGMETRICS '09, pages 49--60,
  New York, NY, USA, 2009. ACM.

\bibitem{karagiannis-blinc}
T.~Karagiannis, K.~Papagiannaki, and M.~Faloutsos.
\newblock Blinc: Multilevel traffic classification in the dark.
\newblock In {\em Proceedings of the 2005 Conference on Applications,
  Technologies, Architectures, and Protocols for Computer Communications},
  SIGCOMM '05, pages 229--240, New York, NY, USA, 2005. ACM.

\bibitem{liberatore2006inferring}
M.~Liberatore and B.~N. Levine.
\newblock Inferring the source of encrypted http connections.
\newblock In {\em Proceedings of the 13th ACM conference on Computer and
  communications security}, pages 255--263. ACM, 2006.

\bibitem{lopez-classifier}
M.~Lopez-Martin, B.~Carro, A.~Sanchez-Esguevillas, and J.~Lloret.
\newblock Network traffic classifier with convolutional and recurrent neural
  networks for internet of things.
\newblock {\em IEEE Access}, 5:18042--18050, 2017.

\bibitem{luchang}
L.~Lu, E.-C. Chang, and M.~C. Chan.
\newblock Website fingerprinting and identification using ordered feature
  sequences.
\newblock In {\em Proceedings of the 15th European Conference on Research in
  Computer Security}, ESORICS'10, pages 199--214, Berlin, Heidelberg, 2010.
  Springer-Verlag.

\bibitem{luo2011}
X.~Luo, P.~Zhou, E.~W.~W. Chan, W.~Lee, R.~K.~C. Chang, and R.~Perdisci.
\newblock {HTTPOS}: Sealing information leaks with browser-side obfuscation of
  encrypted flows.
\newblock In {\em Proceedings of the Network and Distributed System Security
  Symposium (NDSS)}, February 2011.

\bibitem{nguyen-survey-ml}
T.~T. Nguyen and G.~Armitage.
\newblock A survey of techniques for internet traffic classification using
  machine learning.
\newblock {\em Commun. Surveys Tuts.}, 10(4):56--76, Oct. 2008.

\bibitem{homesnitch}
T.~OConnor, R.~Mohamed, M.~Miettinen, W.~Enck, B.~Reaves, and A.-R. Sadeghi.
\newblock {HomeSnitch}: Behavior transparency and control for smart home {IoT}
  devices.
\newblock In {\em Proceedings of the 12th Conference on Security and Privacy in
  Wireless and Mobile Networks}, WiSec '19, pages 128--138, New York, NY, USA,
  2019. ACM.

\bibitem{openwrt}
{OpenWrt/LEDE Project}.
\newblock \url{https://openwrt.org/about}.

\bibitem{panchenko2016website}
A.~Panchenko and F.~Lanze.
\newblock Website fingerprinting at internet scale.
\newblock In {\em NDSS}, 2016.

\bibitem{panchenko2011website}
A.~Panchenko, L.~Niessen, A.~Zinnen, and T.~Engel.
\newblock Website fingerprinting in onion routing based anonymization networks.
\newblock In {\em Proceedings of the 10th annual ACM workshop on Privacy in the
  electronic society}, pages 103--114. ACM, 2011.

\bibitem{perdisci-http-malware}
R.~Perdisci, W.~Lee, and N.~Feamster.
\newblock Behavioral clustering of http-based malware and signature generation
  using malicious network traces.
\newblock In {\em Proceedings of the 7th USENIX Conference on Networked Systems
  Design and Implementation}, NSDI'10, pages 26--26, Berkeley, CA, USA, 2010.
  USENIX Association.

\bibitem{pinheiro2018}
A.~J. {Pinheiro}, J.~M. {Bezerra}, and D.~R. {Campelo}.
\newblock Packet padding for improving privacy in consumer {IoT}.
\newblock In {\em 2018 IEEE Symposium on Computers and Communications (ISCC)},
  pages 00925--00929, June 2018.

\bibitem{ren2019information}
J.~Ren, D.~J. Dubois, D.~Choffnes, A.~M. Mandalari, R.~Kolcun, and H.~Haddadi.
\newblock Information exposure from consumer {IoT} devices: A multidimensional,
  network-informed measurement approach.
\newblock In {\em Proceedings of the Internet Measurement Conference}, pages
  267--279, 2019.

\bibitem{rfctls13}
E.~Rescorla.
\newblock {The Transport Layer Security (TLS) Protocol Version 1.3}.
\newblock {RFC} 8446, {RFC Editor}, August 2018.

\bibitem{tls-rfc}
E.~Rescorla and T.~Dierks.
\newblock {The Transport Layer Security (TLS) Protocol Version 1.2}.
\newblock RFC 5246, Aug. 2008.

\bibitem{saponas}
T.~S. Saponas, J.~Lester, C.~Hartung, S.~Agarwal, and T.~Kohno.
\newblock Devices that tell on you: Privacy trends in consumer ubiquitous
  computing.
\newblock In {\em Proceedings of 16th USENIX Security Symposium on USENIX
  Security Symposium}, SS'07, pages 5:1--5:16, Berkeley, CA, USA, 2007. USENIX
  Association.

\bibitem{sharafaldin2018toward}
I.~Sharafaldin, A.~H. Lashkari, and A.~A. Ghorbani.
\newblock Toward generating a new intrusion detection dataset and intrusion
  traffic characterization.
\newblock 2018.

\bibitem{sivanathan-classifier-new}
A.~Sivanathan, H.~H. Gharakheili, A.~R. Franco~Loi, C.~Wijenayake,
  A.~Vishwanath, and V.~Sivaraman.
\newblock Classifying {IoT} devices in smart environments using network traffic
  characteristics.
\newblock {\em IEEE Transactions on Mobile Computing}, (01):1--1.

\bibitem{sivanathan-classifier}
A.~Sivanathan, D.~Sherratt, H.~H. Gharakheili, A.~Radford, C.~Wijenayake,
  A.~Vishwanath, and V.~Sivaraman.
\newblock Characterizing and classifying {IoT} traffic in smart cities and
  campuses.
\newblock In {\em 2017 IEEE Conference on Computer Communications Workshops
  (INFOCOM WKSHPS)}, pages 559--564, May 2017.

\bibitem{ifttt-news}
{Square, Inc.}
\newblock {What’s going to happen with IFTTT?}
\newblock \url{https://square.github.io/okhttp/}, 2019.

\bibitem{okhttp}
{Stacey Higginbotham}.
\newblock {OkHttp}.
\newblock \url{https://staceyoniot.com/whats-going-to-happen-with-ifttt/},
  2019.

\bibitem{pingpong-software}
R.~Trimananda, J.~Varmarken, A.~Markopoulou, and B.~Demsky.
\newblock Pingpong: Packet-level signatures for smart home devices (software
  and dataset).
\newblock \url{http://plrg.ics.uci.edu/pingpong/}.

\bibitem{devicelist}
UNSW.
\newblock List of smart home devices.
\newblock \url{https://iotanalytics.unsw.edu.au/resources/List_Of_Devices.txt},
  November 2018.

\bibitem{wang2014effective}
T.~Wang, X.~Cai, R.~Nithyanand, R.~Johnson, and I.~Goldberg.
\newblock Effective attacks and provable defenses for website fingerprinting.
\newblock In {\em 23rd $\{$USENIX$\}$ Security Symposium ($\{$USENIX$\}$
  Security 14)}, pages 143--157, 2014.

\bibitem{wright2010}
C.~V. Wright, L.~Ballard, S.~E. Coull, F.~Monrose, and G.~M. Masson.
\newblock Uncovering spoken phrases in encrypted voice over {IP} conversations.
\newblock {\em ACM Transactions on Information and System Security},
  13(4):35:1--35:30, Dec. 2010.

\bibitem{wrightusenix}
C.~V. Wright, L.~Ballard, F.~Monrose, and G.~M. Masson.
\newblock Language identification of encrypted {VoIP} traffic: {Alejandra} y
  {Roberto} or {Alice} and {Bob}?
\newblock In {\em Proceedings of 16th USENIX Security Symposium on USENIX
  Security Symposium}, SS'07, pages 4:1--4:12, Berkeley, CA, USA, 2007. USENIX
  Association.

\bibitem{homonit}
W.~Zhang, Y.~Meng, Y.~Liu, X.~Zhang, Y.~Zhang, and H.~Zhu.
\newblock {HoMonit}: Monitoring smart home apps from encrypted traffic.
\newblock In {\em Proceedings of the 2018 ACM SIGSAC Conference on Computer and
  Communications Security}, CCS '18, pages 1074--1088, New York, NY, USA, 2018.
  ACM.

\end{thebibliography}
}

\newpage
\appendix

\begin{table*}[htb!]
  \centering
  \begin{center}
  { \footnotesize
  \begin{tabular}{| p{11mm} | L{13mm} | c | c | c | c | c | c | c | c | c |}
    \hline
        & & & & & \multicolumn{6}{ c |}{\textbf{Matching (Per 100 Events)}}\\
    \cline{6-11}
        \textbf{Device} & \textbf{Event} & \textbf{Signature} & \textbf{Comm.} & \textbf{Duration (ms)} & \multicolumn{4}{ c |}{\textbf{No Defense}} & \multicolumn{2}{ c |}{\textbf{STP+VPN}} \\
	\cline{6-11}
        & & & & \textbf{Min./Avg./Max.} & \textbf{WAN} & \textbf{FPR} & \textbf{Wi-Fi} & \textbf{FPR} & \textbf{WAN} & \textbf{FPR}\\
        & & & & & \textbf{Snif.} & & \textbf{Snif.} & & \textbf{Snif.} & \\
    \hline
    	\multicolumn{11}{| c |}{\textbf{Plugs}}\\
	\hline
        Amazon & ON & \textbf{S1:} S-[443-445] & \devicecloudcomm & 1,232 / 2,465 / 4,537 & 98 & 0 & 99 & 0 & 99 & 0\\
        plug & & \textbf{S2:} C-1099 S-235 & & & & & & & & \\
	\cline{2-3}
        & OFF & \textbf{S1:} S-[444-446] & & & & & & & & \\
        & & \textbf{S2:} C-1179 S-235 & & & & & & & & \\
        & & \textbf{S3:} C-1514 C-103 S-235 & & & & & & & & \\
	\hline
        WeMo & ON/OFF & \textbf{S1:} PH-259 PH-475 D-246 & \phonedevicecomm & 33 / 42 / 134 & - & - & 100 & 0 & - & - \\
        plug & & & & & & & & & & \\
	\hline
        WeMo & ON/OFF & \textbf{S1:} PH-259 PH-475 D-246 & \phonedevicecomm & 32 / 39 / 97 & - & - & 99 & 0 & - & - \\
        Insight & & & & & & & & & & \\
        plug & & & & & & & & & & \\
	\hline
        TP-Link & ON & \textbf{S1:} C-556 S-1293 & \devicecloudcomm & 75 / 85 / 204 & 99 & 0 & - & - & 98 & 3 \\
    \cline{2-3}
        plug & OFF & \textbf{S1:} C-557 S-[1294-1295] & & & & & & & & \\
    \cline{2-11}
        & ON & \textbf{S1:} PH-112 D-115 & \phonedevicecomm & 225 / 325 / 3,328 & - & - & 99 & 0 & - & - \\
        & & \textbf{S2:} C-556 S-1293 & \& & & & & & & & \\
	\cline{2-3}
	     & ON & \textbf{S1:} PH-112 D-115 & \devicecloudcomm & & & & & & & \\
        & & \textbf{S2:} C-557 S-[1294-1295] & & & & & & & & \\
	\hline
        D-Link plug & ON/OFF & \textbf{S1:} S-91 S-1227 C-784 & \devicecloudcomm & 4 / 1,194 / 8,060 & 95 & 0 & 95 & 0 & 95 & 0 \\
        & & \textbf{S2:} C-1052 S-647 & & & & & & & & \\
    \cline{2-11}
    	& ON & \textbf{S1:} C-[1109-1123] S-613 & \phonecloudcomm & 35 / 41 / 176 & 98 & 0 & 98 & 0 & 98 & 0\\
    \cline{2-3}
        & OFF & \textbf{S1:} C-[1110-1124] S-613 & & & & & & & & \\
    \hline
        Smart- & ON & \textbf{S1:} C-699 S-511 & \phonecloudcomm & 335 / 537 / 2,223 & 92 & 0 & 92 & 0 & 92 & 0\\
        Things & & \textbf{S2:} S-777 C-136 & & & & & & & & \\
	\cline{2-3}
        plug & OFF & \textbf{S1:} C-700 S-511 & & & & & & & & \\
        & & \textbf{S2:} S-780 C-136 & & & & & & & & \\
    \hline
    	\multicolumn{11}{| c |}{\textbf{Light Bulbs}}\\
	\hline
        Sengled & ON & \textbf{S1:} S-[217-218] C-[209-210] & \devicecloudcomm & 4,304 / 6,238 / 8,145 & 97 & 0 & - & - & 97 & 0\\
        light & & \textbf{S2:} C-430 & & & & & & & & \\
        bulb & & \textbf{S3:} C-466 & & & & & & & & \\
	\cline{2-3}
        & OFF & \textbf{S1:} S-[217-218] C-[209-210] & & & & & & & & \\
        & & \textbf{S2:} C-430 & & & & & & & & \\
        & & \textbf{S3:} C-465 & & & & & & & & \\
	\cline{2-11}
        & ON & \textbf{S1:} C-211 S-1063 & \phonecloudcomm & 4,375 / 6,356 / 9,132 & 93 & 0 & 97 & 0 & 96 & 1\\
        & & \textbf{S2:} S-1277 & & & & & & & & \\
	\cline{2-3}
        & OFF & \textbf{S1:} C-211 S-1063 S-1276 & & & & & & & & \\
	\cline{2-11}
        & Intensity & \textbf{S1:} S-[216-220] & \devicecloudcomm & 16 / 74 / 824 & 99 & 2 & - & - & 99 & 5\\
        & & C-[208-210] & & & & & & & & \\
	\cline{2-11}
        & Intensity & \textbf{S1:} C-[215-217] & \phonecloudcomm & 3,916 / 5,573 / 7,171 & 99 & 0 & 99 & 0 & 98 & 2 \\
        & & S-[1275-1277] & & & & & & & & \\
	\hline
        Hue light & ON & \textbf{S1:} C-364 & \devicecloudcomm & 11,019 / 12,787 / & - & - & - & - & - & -\\
        bulb & & \textbf{S2:} D-88 & \& & 14,353 & & & & & & \\
	\cline{2-3}
        & OFF & \textbf{S1:} C-365 & \phonedevicecomm & & & & & & & \\
		& & \textbf{S2:} D-88 & & & & & & & & \\
	\hline
        TP-Link & ON & \textbf{S1:} PH-198 D-227 & \phonedevicecomm & 8 / 77 / 148 & - & - & 100 & 4 & - & - \\
	\cline{2-3}
        light & OFF & \textbf{S1:} PH-198 D-244 & & & & & & & & \\
    \cline{2-11}
        bulb & Intensity & \textbf{S1:} PH-[240-242] D-[287-289] & \phonedevicecomm & 7 / 84 / 212 & - & - & 100 & 0 & - & - \\
    \cline{2-11}
        & Color & \textbf{S1:} PH-317 D-287 & \phonedevicecomm & 6 / 89 / 174 & - & - & 100 & 0 & - & - \\
    \hline
    	\multicolumn{11}{| c |}{\textbf{Thermostats}}\\
	\hline
    	Nest & Fan ON & \textbf{S1:} C-[891-894] S-[830-834] & \phonecloudcomm & 91 / 111 / 1,072 & 93 & 0 & 93 & 1 & 93 & 2 \\
    \cline{2-3}
    	thermostat & Fan OFF & \textbf{S1:} C-[858-860] S-[829-834] & & & & & & & & \\
   	\hline
        Ecobee & HVAC Auto & \textbf{S1:} S-1300 C-640 & \phonecloudcomm & 121 / 229 / 667 & 100 & 0 & 99 & 0 & 99 & 0\\
	\cline{2-3}
        thermostat & HVAC OFF & \textbf{S1:} C-1299 C-640 & & & & & & & & \\
	\cline{2-11}
        & Fan ON & \textbf{S1:} S-1387 C-640 & \phonecloudcomm & 117 / 232 / 1,776 & 100 & 0 & 100 & 0 & 100 & 1\\
	\cline{2-3}
        & Fan Auto & \textbf{S1:} C-1389 C-640 & & & & & & & & \\
    \hline
    	\multicolumn{11}{| c |}{\textbf{Sprinklers}}\\
	\hline
        Rachio & Quick Run & \textbf{S1:} S-267 C-155 & \devicecloudcomm & 1,972 / 2,180 / 2,450 & 100 & 0 & 100 & 0 & 100 & 1\\
	\cline{2-3}
        sprinkler & Stop & \textbf{S1:} C-496 C-155 C-395 & & & & & & & & \\
	\cline{2-11}
        & Standby/ Active & \textbf{S1:} S-299 C-155 C-395 & \devicecloudcomm & 276 / 690 / 2,538 & 100 & 0 & 100 & 0 & 100 & 0\\
    \hline
        Blossom & Quick Run & \textbf{S1:} C-326 & \devicecloudcomm & 701 / 3,470 / 8,431 & 96 & 0 & 96 & 0 & 96 & 3\\
        sprinkler & & \textbf{S2:} C-177 S-505 & & & & & & & & \\
	\cline{2-3}
        & Stop & \textbf{S1:} C-326 & & & & & & & & \\
        & & \textbf{S2:} C-177 S-458 & & & & & & & & \\
        & & \textbf{S3:} C-238 C-56 S-388 & & & & & & & & \\
    \cline{2-11}
        & Quick Run & \textbf{S1:} C-649 S-459 C-574 S-507 & \phonecloudcomm & 70 / 956 / 3,337 & 93 & 0 & 93 & 0 & 93 & 0\\
        & & \textbf{S2:} S-[135-139] & & & & & & & & \\
	\cline{2-3}
        & Stop & \textbf{S1:} C-617 S-431 & & & & & & & & \\
    \hline
  \end{tabular}
  }
  \end{center}
\end{table*}

\begin{table*}[htb!]
  \centering
  \begin{center}
  { \footnotesize
  \begin{tabular}{| p{14mm} | L{13mm} | c | c | c | c | c | c | c | c | c |}
    \hline
        & & & & & \multicolumn{6}{ c |}{\textbf{Matching (Per 100 Events)}}\\
    \cline{6-11}
        \textbf{Device} & \textbf{Event} & \textbf{Signature} & \textbf{Comm.} & \textbf{Duration (ms)} & \multicolumn{4}{ c |}{\textbf{No Defense}} & \multicolumn{2}{ c |}{\textbf{STP+VPN}} \\
	\cline{6-11}
        & & & & \textbf{Min./Avg./Max.} & \textbf{WAN} & \textbf{FPR} & \textbf{Wi-Fi} & \textbf{FPR} & \textbf{WAN} & \textbf{FPR}\\
        & & & & & \textbf{Snif.} & & \textbf{Snif.} & & \textbf{Snif.} & \\
	\hline
        & Hibernate & \textbf{S1:} C-621 S-493 & \phonecloudcomm & 121 / 494 / 1,798 & 95 & 0 & 93 & 0 & 93 & 1\\
	\cline{2-3}
        & Active & \textbf{S1:} C-622 S-494 & & & & & & & & \\
        & & \textbf{S2:} S-599 C-566 S-554 C-566 & & & & & & & & \\
		\hline
    	\multicolumn{11}{| c |}{\textbf{Home Security Devices}}\\

		\hline
        Ring alarm & Arm & \textbf{S1:} S-99 S-254 C-99 & \devicecloudcomm & 275 / 410 / 605 & 98 & 0 & 95 & 0 & 95 & 0\\
        & & S-[181-183] C-99 & & & & & & & & \\
	\cline{2-3}
        & Disarm & \textbf{S1:} S-99 S-255 C-99 & & & & & & & & \\
        & & S-[181-183] C-99 & & & & & & & & \\
    \hline
        Arlo & Stream ON & \textbf{S1:} C-[338-339] S-[326-329] & \phonecloudcomm & 46 / 78 / 194 & 99 & 2 & 98 & 3 & 97 & 4\\
        camera & & C-[364-365] S-[1061-1070] & & & & & & & & \\
        & & \textbf{S2:} C-[271-273] S-[499-505] & & & & & & & & \\
	\cline{2-3}
        & Stream OFF & \textbf{S1:} C-[445-449] S-442 & & & & & & & & \\
	\hline
        D-Link & ON & \textbf{S1:} C-1076 S-593 & \phonecloudcomm & 36 / 37 / 65 & 100 & 0 & 98 & 0 & 97 & 0 \\
	\cline{2-3}
        siren & OFF & \textbf{S1:} C-1023 S-613 & & & & & & & & \\
	\hline
        Kwikset & Lock & \textbf{S1:} C-699 S-511 & \phonecloudcomm & 173 / 395 / 2,874 & 100 & 0 & 100 & 0 & 100 & 0\\
        door & & \textbf{S2:} S-639 C-136 & & & & & & & & \\
	\cline{2-3}
        lock & Unlock & \textbf{S1:} C-701 S-511 & & & & & & & & \\
        & & \textbf{S2:} S-647 C-136 & & & & & & & & \\
    \hline
    	\multicolumn{11}{| c |}{\textbf{Others}}\\
	\hline
        Roomba & Clean & \textbf{S1:} S-[1014-1015] C-105 & \phonecloudcomm & 123 / 2,038 / 5,418 & 91 & 0 & 94 & 0 & 94 & 1\\
        robot & & S-432 C-105 & & & & & & & & \\
	\cline{2-3}
        & Back-to- & \textbf{S1:} S-440 C-105 & & & & & & & & \\
        & station & S-[1018-1024] C-105 & & & & & & & & \\
    \hline
    	\multicolumn{5}{| r |}{Average} & 97.05 & 0.18 & 97.48 & 0.32 & 96.77 & 1.09 \\
    \hline
  \end{tabular}
  }
  \end{center}
	\caption{Smart home devices found to exhibit \phonecloudcomm, \devicecloudcomm, and \phonedevicecomm signatures.
  	Prefix PH indicates Phone-to-device direction and prefix D indicates Device-to-phone direction in Signature column.
  	\label{tab:summary-signatures-cr}}
	
	\vspace{-1em}
\end{table*}

\subsection{Detection\label{sect:system-detection-cr}}

\tool's detection component determines if a packet-level signature is present in a captured network trace, 
\athina{or real-time network traffic.}
The implementation differs slightly for the two adversaries described in Section~\ref{sect:threat-model}.

\myparagraph{Layer-2 Information.}
The \wifisniffer has the disadvantage of only having access to layer-2 header information due to WPA2 encryption.
This means that it cannot reconstruct TCP connections, but can only separate traffic into flows of packets exchanged 
between pairs of MAC addresses, referred to as \emph{layer-2 flows}.
On layer 2, the encryption added by WPA2 does not pad packet lengths. 
Thus, our signatures, extracted from TCP/IP traffic, can be directly mapped to layer 2 if the IEEE 802.11 radiotap header, 
frame header, the AES-CCMP IV and key identifier, 
and FCS are accounted for. In our testbed, these consistently add 80 bytes to the packet length.

\myparagraph{Packet Sequence Matching.}
A state machine is maintained for each packet sequence of the signature for each flow, \ie TCP connection for the \wansniffer or 
layer-2 flow for the 
\wifisniffer.
Each packet in the stream of packets is presented to the state machines associated with the flow that the packet pertains to.
If the packet's length and direction match that of the packet for the next state, the state machine advances and records the 
packet.
The detection algorithm operates differently for layer-2 and layer-3 detections when the packet does not match the expected next packet.
For layer-2 detection (\wifisniffer), the packet is simply ignored, and the state machine remains in the same state. 
\majorrevision{Out-of-order packets can in theory cause failures for our current layer-2 signature matching implementation.  In practice, out-of-order packets occur rarely enough that they are unlikely to be a significant issue.  There are other mitigating factors that further lower the likelihood of out-of-order packets posing a problem.  Signatures that strictly follow the pingpong pattern do not provide an opportunity for their packets to be reordered. Out-of-order packets often result in TCP ACKs in the other direction that would cause retransmissions. Thus, another event packet would be seen right after the first one and the first one would be ignored.}
For layer-3 detection (\wansniffer), the packet causes the state machine to discard any partial match---layer-3 detection 
does not deal with interleaving packets as it considers individual TCP connections and can filter out TCP retransmissions.
When a state machine matches its first packet and advances to the next state, a new state machine is created for the same packet sequence, but in the initial state.
This is to ensure that the state machine starts at the correct first packet,
\eg when a packet of that length appears in other traffic.
To bound the number of active state machines, and to minimize the number of false positives resulting from retransmissions, any state machine that advances from state $s$ to state $s+1$ replaces any existing state machine in state $s+1$ \emph{iff} the last packet of the newly advanced state
machine has a later timestamp than that of the existing state machine.
Once a state machine reaches its terminal state, the set of recorded packets is reported as a sequence match.

\myparagraph{Matching Strategies.}
For every packet in a sequence, there are two possible matching strategies: 
\emph{exact} and \emph{range-based} matching.
In exact matching, the state machines only consider exactly those packet lengths that were observed during training as valid.
In the range-based matching strategy, the state machines allow the packet lengths to lie between the minimum and maximum packet lengths (plus a small delta) observed during training.
As such, range-based matching attempts to accommodate packet sequences that have slight variations where all permutations may not have been observed during training.
For range-based matching, the lower and upper bounds for each packet of a packet sequence are derived from the core points of the packet pair clustering (see Section~\ref{sect:clustering}).
$\epsilon$~is then applied to these bounds analogous to the clustering technique used in the DBSCAN
\majorrevision{
algorithm---we, therefore, consistently use the same $\epsilon=10$.
}
For example, for $\epsilon=10$ and core points \texttt{<C-338, S-541>} and \texttt{<C-339, S-542>}, a state machine that uses range-based matching will consider client-to-server packets with lengths in \texttt{[328, 349]}, and server-to-client 
packets with lengths in \texttt{[531,552]} as valid.

Exact matching is used when no variations in packet lengths were observed during training, and range-based matching is used if variations in packet lengths were observed during training.
However, range-based matching is not performed when the signature only consists of 2 packets and/or 
there is an overlap between the signatures that represent different types of \events (\eg the \dlinkplug's signatures for ON and OFF in 
Table~\ref{tab:summary-signatures}) as range-based matching for 2-packet signatures has a high risk of generating
many false positives.

\myparagraph{Declaring a Signature Match.}
A sequence match does not necessarily mean that the full signature
has been matched. Some signatures are comprised of multiple packet
sequences and all of them have to be matched
(\eg \arlocamera, see Section~\ref{sect:system-training}). 
Sequence matches are therefore
reported to a secondary module that verifies if the required temporal constraints are
in place, namely that the sequence match for packet sequence set $i$ occurs
\emph{before} the sequence match for packet sequence set $i+1$ \emph{and}
that the time between the first packet of the sequence match corresponding 
to packet sequence set 1 and the last packet of the sequence match
corresponding to packet sequence set $k$ (for a signature with $k$ packet
sequence sets) is below a threshold.
A signature match is declared when all matching packet sequence sets occur within
the duration $\lceil t + 0.1t \rceil$ with $t$ being the maximum observed signature 
duration \majorrevision{(see Table~\ref{tab:summary-signatures})}.

\myparagraph{Simultaneous Events.}
Finally, during signature matching, \tools algorithm first separates incoming packets into
different sets on a per device basis: individual TCP connections in the WAN sniffer matching, 
or individual substreams based on source/destination MAC addresses in the Wi-Fi sniffer
matching. 
Since the devices we have seen only generate \events sequentially, sequences that
correspond to a certain \event will also appear sequentially in the network trace: there
will never be an overlap of \events in a single device.
If two devices or more generate \events simultaneously, the corresponding sequences will
be either in separate TCP connections for the \wansniffer adversary or separate substreams for the \wifisniffer adversary:
there will never be overlap of \events generated by different devices.
This also implies that changing the amount and types of background traffic 
(e.g., video or audio streaming) does not affect our signatures---other traffic will be in
different TCP connections/substreams.

\begin{table*}[htb!]
  \centering
  \majorrevision{
  \begin{center}
  { \footnotesize
  \begin{tabular}{| p{25mm} | p{20mm} | c | c | c | c | c | c |}
    \hline
        \textbf{Device} & \textbf{Event} & \textbf{\devicecloudcomm} Signature & \textbf{Duration (ms)} & \multicolumn{4}{ c |}{\textbf{Matching (Per 100 Events)}}\\
    \cline{5-8}
        & & & \textbf{Min./Avg./Max.} & \textbf{WAN} & \textbf{FPR} & \textbf{Wi-Fi} & \textbf{FPR}\\
        & & & & \textbf{Sniffer} & & \textbf{Sniffer} & \\
	\hline
		\multicolumn{8}{| c |}{\textbf{Plugs}}\\
	\hline
        \wemoplug & ON/OFF & \textbf{S1:} S-146 & 226 / 345 / 2,236 & 100 & 0 & 100 & 0\\
        & & \textbf{S2:} C-210 S-134 S-286 C-294 & & & & &\\
	\hline
        \wemoinsightplug & ON & \textbf{S1:} S-146 & 216 / 253 / 473 & 99 & 0 & 94 & 0\\
        & & \textbf{S2:} C-210 S-134 S-286 C-294 & & & & &\\
    \cline{2-3}
        & OFF & \textbf{S1:} S-146 & & & & &\\
        & & \textbf{S2:} C-210 S-134 S-350 C-294 & & & & &\\
	\hline
        \tplinkplug & ON & \textbf{S1:} C-592 S-1234 S-100 & 71 / 76 / 139 & 100 & 0 & 100 & 0\\
    \cline{2-3}
        & OFF & \textbf{S1:} C-593 S-1235 S-100 & & & & &\\
	\hline
        \dlinkplug & ON/OFF & \textbf{S1:} C-256 & 269 / 750 / 6,364 & 93 & 1 & 93 & 1\\
        & & \textbf{S2:} C-1020 S-647 & & & & & \\
    \hline
		\multicolumn{8}{| c |}{\textbf{Light Bulbs}}\\
	\hline
        \huebulb 
         & ON & \textbf{S1:} S-[227-229] C-[857-859] C-365 & 37 / 44 / 76 & 99 & 1 & - & -\\
	\cline{2-3}
        & OFF & \textbf{S1:} S-[227-230] C-[857-860] C-366 & & & & &\\
    \cline{2-8}
        & Intensity & \textbf{S1:} S-[237-240] C-[895-899] & 36 / 40 / 96 & 97 & 0 & - & -\\
        & & \textbf{S2:} C-[378-379] & & & & &\\
	\hline
        \tplinkbulb & ON & \textbf{S1:} S-[348-349] C-[399-400] & 11 / 15 / 78 & 100 & 0 & 100 & 0\\
	\cline{2-3}
        & OFF & \textbf{S1:} S-[348-349] C-[418-419] & & & & &\\
    \cline{2-8}
        & Intensity & \textbf{S1:} S-[438-442] C-[396-400] & 12 / 16 / 53 & 100 & 0 & 99 & 0\\
    \cline{2-8}
        & Color & \textbf{S1:} S-[386-388] C-[397-399] & 12 / 17 / 60 & 99 & 0 & 97 & 0\\
    \hline
		\multicolumn{8}{| c |}{\textbf{Others}}\\
	\hline
        \rachiosprinkler & Quick Run & \textbf{S1:} S-267 C-155 & 1,661 / 2467 / 4,677 & 95 & 3 & 95 & 5\\
	\cline{2-3}
        & Stop & \textbf{S1:} C-661 & & & & &\\
				& & \textbf{S2:} C-155 & & & & &\\
	\hline
        \arlocamera & Start Recording & \textbf{S1:} C-704 S-215 & 156 / 159 / 195 & 100 & 0 & 99 & 0\\
    \hline
        \dlinksiren & ON & \textbf{S1:} S-[989-1005] C-616 & 162 / 181 / 281 & 99 & 1 & 98 & 1\\
        & & \textbf{S2:} C-216 & & & & &\\
    \hline
    	\multicolumn{4}{| r |}{Average} & 98.4 & 0.5 & 97.5 & 0.7 \\
    \hline
  \end{tabular}
  }
  \end{center}
  \caption{\devicecloudcomm signature  in 9 devices triggered by IFTTT home automation.\label{tab:home-automation}}
  }
\end{table*}

\begin{table*}[htb!]
  \centering
  \majorrevision{
  \begin{center}
  { \footnotesize
  \begin{tabular}{| p{20mm} | p{15mm} | c | c | c |}
    \hline
        \textbf{Device} & \textbf{Event} & \multicolumn{3}{ c |}{\textbf{\devicecloudcomm Signatures}}\\
    \cline{3-5}
        & & \textbf{\localphone} & \textbf{\remotephone} & \textbf{\ifttt}\\
	\hline
        \tplinkplug & ON & \textbf{S1:} C-592 S-1234 S-100 & \textbf{S1:} C-592 S-1234 S-100 & \textbf{S1:} C-592 S-1234 S-100 \\
    \cline{2-5}
        & OFF & \textbf{S1:} C-593 S-1235 S-100 & \textbf{S1:} C-593 S-1235 S-100 & \textbf{S1:} C-593 S-1235 S-100 \\
	\hline
        \dlinkplug & ON/OFF & \textbf{S1:} S-91 S-1227 C-784 & \textbf{S1:} S-91 S-1227 C-784 & \textbf{S1:} C-256\\
        & & \textbf{S2:} C-1052 S-647 & \textbf{S2:} C-1052 S-647 & \textbf{S2:} C-1020 S-647 \\
	\hline
        \rachiosprinkler & Quick Run & \textbf{S1:} S-267 C-155 & \textbf{S1:} S-267 C-155 C-837 C-448 & \textbf{S1:} S-267 C-155 \\
    \cline{2-5}
        & Stop & \textbf{S1:} C-496 C-155 C-395 & \textbf{S1:} S-219 S-235 C-171 C-661 C-496 C-155 C-395 & \textbf{S1:} C-661\\
        & & & & \textbf{S2:} C-155\\
	\hline
  \end{tabular}
  }
  \end{center}
  \caption{Comparison of \devicecloudcomm signatures for three devices (\tplinkplug, \dlinkplug, and \rachiosprinkler) triggered in three different ways: via (i) a local phone, (ii) a remote phone, and (iii) IFTTT home automation.\label{tab:device-cloud-comparison}}
  }
\end{table*}

\subsection{Events Triggered Remotely}
\label{sect:home-automation-cr}

Our main dataset, collected using our testbed (see Section~\ref{sect:training-phase}), contains \events triggered by a smartphone that is part of the local network. 
 However, smart home devices can also be controlled remotely, using home automation frameworks or a remote smartphone. In this section, we show that even with such remote triggers, the \devicecloudcomm communication exhibits similar signatures, which can be extracted using \tool{}.

\myparagraph{Home Automation Experiment (\ifttt).}
In this section, we trigger events using the \ifttt (If-This-Then-That) home automation platform~\cite{ifttt}, which supports most of our devices. 
\ifttt is one of the most popular web-based home automation 
frameworks: in 2019 it provides around 700 services for 17 million users~\cite{ifttt-news}.
\ifttt allows users to set up automation rules that work in a \emph{Trigger-Action} fashion.
Each automation rule typically connects a trigger (\ie \emph{This}) to an action (\ie \emph{That}).
The occurrence of the trigger causes the action to be performed by the framework.
For example, ``if motion is detected, then turn on the smart plug''~\cite{ifttt-motion}.

In this experiment, we use \ifttt to trigger \events. We integrate \ifttt into our existing infrastructure for triggering 
device events via Android widget button presses.
For each event type, we develop an \ifttt rule that triggers the \event using an 
Android button widget.
For example, we set up a button widget to toggle ON 
the \tplinkplug: ``if the widget button is pressed, then turn on the smart plug''.
Then, we install the \ifttt app on our smartphone, log in to the app using our \ifttt account, 
and add the button widget onto the home screen of the smartphone.
Then, instead of using the official Android app, we use the button widget 
to trigger \events from the smartphone. 
The smartphone is connected to a different network than the smart home testbed network to simulate controlling the devices from a remote location.

For every device: (1) we use \ifttt to trigger events to collect a new training  dataset,  and we run \tool to extract packet-level signatures from the \devicecloudcomm  communication in this dataset; (2) we collect a new smart home testbed dataset (as in Section~\ref{sect:detection-phase}, but using the \ifttt widget to trigger events), which we then use for testing, \ie to detect \events by matching on the \ifttt signatures extracted from the aforementioned training dataset.


\mysubparagraph{Signatures Found in \devicecloudcomm Communication.}
\ifttt provides support for 13 out of our 18 devices: no support was  provided at the time of the experiment 
for the \amazonplug, \blossomsprinkler, \roombarobot, \ringalarm, and \nestthermostat.
The main finding is that, from the supported 13 devices, \tool{} successfully extracts \devicecloudcomm signatures 
for nine devices and 12 \event types, which are summarized in 
Table~\ref{tab:home-automation}.\footnote{\majorrevision{\tool did not extract \devicecloudcomm signatures from 4 devices:
the \sengledbulb, \ecobeethermostat, \smartthingsplug, and \kwiksetdoorlock. 
For the \ecobeethermostat, \smartthingsplug, and \kwiksetdoorlock. \tool extracted
signatures from the \phonecloudcomm communication (not from the \devicecloudcomm communication) 
in the previous experiment (see Table~\ref{tab:summary-signatures-cr}).
For the \sengledbulb, the device was recently forced to update its firmware---\tool in its current state did
not extract signatures in the \devicecloudcomm communication anymore although there is 
potentially a new signature.
}}
Three out of the nine supported devices (the \tplinkplug, the \dlinkplug, and the \rachiosprinkler) already 
have \devicecloudcomm signatures when triggered by a local phone: the phone is connected
to the smart home testbed, where the device is also connected to (see Table~\ref{tab:summary-signatures-cr}).
Interestingly, six out of the nine supported devices  have \devicecloudcomm signatures when 
triggered via \ifttt, but did not have \devicecloudcomm signatures when triggered by a local phone. 

\myparagraph{Comparison of \devicecloudcomm Signatures.}
Having answered the main question (\ie that there are indeed \devicecloudcomm signatures even when devices are triggered by IFTTT), a secondary question is whether the  signature varies depending on the way the device is triggered. 
To answer this question, we consider the \tplinkplug, the \dlinkplug, and the \rachiosprinkler, and we  trigger events in three different ways:
\squishcount
\item \localphone: signatures are extracted from the previous 
experiment (Table~\ref{tab:summary-signatures-cr}), using the vendor's official
Android application to trigger \events.  The phone is connected 
to the smart home testbed network.
\item \remotephone: signatures are extracted from training datasets we collect using a remote phone setting (without using \ifttt). We  use the vendor's official Android 
application to trigger \events for each device. We connect the phone to a different
network than the smart home testbed network.
\item \ifttt: signatures are extracted from a training dataset collected with the \ifttt home automation experiment described in this section.
\countend
 
Table~\ref{tab:device-cloud-comparison} lists all the \devicecloudcomm signatures  we extract.
We can see that the majority of \devicecloudcomm signatures are the same or very similar across the \ifttt, \localphone, and \remotephone experiments.
For the \tplinkplug, the \devicecloudcomm signatures from the three experiments are the same or
similar (same packet sequences within 1B).\footnote{\majorrevision{We repeated our previous experiment and found that
the \devicecloudcomm signatures presented in Table~\ref{tab:summary-signatures-cr} have evolved over time and since the original experiment.}}
For the \dlinkplug,  the \localphone and the \remotephone \devicecloudcomm signatures are  the same,
but the \ifttt \devicecloudcomm signatures are partially similar to the 
\localphone and \remotephone \devicecloudcomm signatures.
For the \rachiosprinkler, the \devicecloudcomm signatures from the three experiments
are subsets of one another. 

\begin{table*}[htb!]
  \centering
  \majorrevision{
  \begin{center}
  { \footnotesize
  \begin{tabular}{| p{30mm} | p{10mm} | p{10mm} | c | c |}
    \hline
        \textbf{Device} & \textbf{Model} & \textbf{Event} & \textbf{Signature} & \textbf{Duration (ms)}\\
        & & & & \textbf{Min./Avg./Max.}\\
	\hline
		\multicolumn{5}{| c |}{\textbf{Existing TP-Link Devices}}\\
	\hline
        \tplinkplug & HS-110 & ON & $\ast$\textbf{S1:} PH-172 D-115 & 406 / 743 / 10,667\\
        & & & \textbf{S2:} C-592 S-1234 S-100 & \\
    \cline{3-4}
        & & OFF & $\ast$\textbf{S1:} PH-172 D-115 & \\
        & & & \textbf{S2:} C-593 S-1235 S-100 & \\
	\hline
        \tplinkbulb & LB-130 & ON & $\ast$\textbf{S1:} PH-258 D-288 & 8 / 77 / 148 \\
    \cline{3-4}
        & & OFF & $\ast$\textbf{S1:}  PH-258 D-305 & \\
    \cline{3-5}
        & & Intensity & \textbf{S1:} PH-[240-242] D-[287-289] & 7 / 84 / 212\\
    \cline{3-5}
        & & Color & \textbf{S1:} S1: PH-317 D-287 & 6 / 89 / 174\\
	\hline
		\multicolumn{5}{| c |}{\textbf{Newly Added TP-Link Devices}}\\
	\hline
        \tplinktwooutletplug & HS-107 & ON & \textbf{S1:} PH-219 D-103 & 1,083 / 1593 / 2,207\\
        & & & \textbf{S2:} C-300 C-710 S-1412 S-88 & \\
    \cline{3-4}
        & & OFF & \textbf{S1:} PH-219 D-103 & \\
        & & & \textbf{S2:} C-300 C-711 S-1413 S-88 & \\
	\hline
        \tplinkpowerstrip & HS-300 & ON & \textbf{S1:} PH-219 D-103 & 976 / 1,537 / 4,974\\
        & & & \textbf{S2:} C-301 C-1412 S-[1405-1406] S-88 & \\
    \cline{3-4}
        & & OFF & \textbf{S1:} PH-219 D-103 & \\
        & & & \textbf{S2:} C-301 C-1413 S-[1406-1407] S-88 & \\
	\hline
        \tplinkwhitebulb & KL-110 & ON & \textbf{S1:} S-[414-415] C-[331-332] & 1,892 / 2,021 / 2,157\\
        & & & \textbf{S2:} C-648 S-[1279-1280] S-88 & \\
    \cline{3-4}
        & & OFF & \textbf{S1:} S-[414-415] C-[350-351] & \\
        & & & \textbf{S2:} C-649 S-[1280-1281] S-88 & \\
    \cline{3-5}
        & & Intensity & \textbf{S1:} S-[479-483] C-[329-332] & 2,418 / 2,540 / 3,610\\
        & & & \textbf{S2:} C-[654-656] S-[1285-1288] S-88 & \\
	\hline
        \tplinkcamera & KC-100 & ON & \textbf{S1:} PH-256 D-162 PH-624 D-256 PH-72 D-111 PH-608 D-371 PH-97 &  804 / 1,105 / 1,739\\
        & & & \textbf{S2:} C-1288 S-[1161-1162] S-100 & \\
    \cline{3-4}
        & & OFF & \textbf{S1:} PH-256 D-162 PH-624 D-256 PH-72 D-111 PH-614 D-371 PH-97 & \\
        & & & \textbf{S2:} C-1289 S-[1162-1163] S-100 & \\
	\hline
  \end{tabular}
  }
  \end{center}
  \caption{Signatures extracted from different TP-Link devices.
	$\ast$These are the latest signatures for the \tplinkplug and \tplinkbulb (per December 2019). 
	\label{tab:same-vendor}}
  }
\end{table*}

{

\subsection{Traffic Shaping and Injection}
\label{sect:defenses-packet-injection-cr}
Stochastic traffic padding (STP)\footnote{\majorrevision{It is worth clarifying that despite its name, STP does not perform packet padding. It performs traffic shaping and injection (referred to as ``traffic padding'' in~\cite{princeton-stp}) of fake traffic that resembles the real IoT traffic.}}  is a state-of-the-art defense for passive inference attack on smart home devices~\cite{princeton-stp}.\footnote{\majorrevision{Other examples of traffic injection include  \cite{panchenko2011website}  that injects fake/spurious HTTP requests to defend against website fingerprinting.}}
STP shapes real network traffic generated by IoT \events and injects fake \event traffic
randomly into upload and download traffic to imitate volume-based signatures.
We contacted the authors of ~\cite{princeton-stp}, but they did not share their STP implementation.
Therefore, we simulated the VPN-based STP implementation, and performed a simple but conservative test. 
We used OpenVPN to replay packets from our pre-recorded smart home device events. Alongside 100 real events, we injected 100 dummy STP \events of the same type 
distributed evenly and randomly throughout the experiment. Our experiments with OpenVPN reveal that it consistently adds a header of 52 bytes for client-to-server packets and 49 bytes for  server-to-client packets: thus our signatures remain intact. However, all traffic is now combined into one flow between two endpoints, and this could potentially increase the FPs. 

\majorrevision{
	In our STP experiments, \tool performs well and STP has very little effect on
	our signatures: it does not generate many FPs: 
	the average recall remains around 97\% and the FPR increases to 1.09 per 100 dummy \events of the same \event type: the FPR increase is minimal.
	Table~\ref{tab:summary-signatures-cr} reports additional false positives for 3 
	devices: \arlocamera, \nestthermostat, and \tplinkbulb. 
	Further inspection revealed that these FPs were caused only by 2-packet signatures, while longer 
	signatures were more resilient.}
The small increase occurs for two reasons: (1) the VPN tunnel combines all traffic into one flow, and (2)
the dummy \event packets are sometimes coincidentally sent from both endpoints simultaneously, allowing 
the request and reply packets to be in the same signature duration window (see Section~\ref{sect:parameter-sensitivity}).
Thus, a VPN-based STP implementation is not effective in defending against our \tool{}; this is expected as STP was designed to defend against volume-based signatures, while our signatures consist of sequences of packet lengths, which survive both traffic shaping and injection.
Furthermore, this defense is currently only applicable against the \wansniffer adversary, while the \wifisniffer   is not affected by this router-based implementations of STP.

%
%

}

\end{document}